\documentclass[sigplan,10pt,nonacm]{acmart}
\renewcommand\footnotetextcopyrightpermission[1]{}
\usepackage{amsmath}
\usepackage{mathtools}
\usepackage{bm}

\usepackage{amsthm}
\makeatletter
\newtheoremstyle{nonbolddef}
  {\topsep}    %
  {\topsep}    %
  {\normalfont} %
  {}           %
  {\itshape} %
  {.}          %
  {.5em}       %
  {}           %
\makeatother
\theoremstyle{nonbolddef}

\newtheorem{definition}{Definition}

\usepackage{tikz}
\usepackage{multirow}
\usepackage{pgfplots}
\usepackage{xspace}
\usepackage{xcolor}
\usetikzlibrary{patterns}
\usetikzlibrary{calc,positioning,fit}
\usetikzlibrary{shapes, matrix}
\usetikzlibrary{arrows.meta}
\usetikzlibrary{decorations.markings}

\usepgfplotslibrary{groupplots,units}

\usepackage{graphicx}
\graphicspath{{figures/}}
\usepackage[linesnumbered, ruled, vlined]{algorithm2e}
\usepackage{hyperref}
\hypersetup{
    colorlinks=true,
    linkcolor=blue,      %
    citecolor=blue,      %
    filecolor=blue,
    urlcolor=black        %
}

\usepackage[capitalise]{cleveref}
\crefname{section}{§}{§§}
\Crefname{section}{§}{§§}
\crefformat{section}{\S#2#1#3}
\crefname{figure}{Fig.}{Figs.}
\Crefname{figure}{Fig.}{Figs.}
\crefname{table}{Tab.}{Tabs.}
\Crefname{table}{Tab.}{Tabs.}

\usepackage{url}

\usepackage{breakurl}

\usepackage{subcaption}
\usepackage{listings}

\usepackage{soul}
\sethlcolor{lightgray}
\usepackage[export]{adjustbox}
\usepackage{tabularx}
\usepackage{makecell}
\usepackage{booktabs}
\usepackage{float}
\usepackage[normalem]{ulem}

\usepackage{pgfplotstable}
\pgfplotsset{compat=1.18}

\usepgfplotslibrary{groupplots,units,fillbetween,statistics}
\usepackage{siunitx}
\usepackage{pifont}
\usepackage{colortbl}

\usepackage[shortlabels]{enumitem}
\setlist[enumerate]{leftmargin=1cm,topsep=0.5mm}
\setlist[itemize]{leftmargin=0.5cm,topsep=0.5mm,noitemsep}
\setenumerate{label={\arabic*.},noitemsep,topsep=2pt,leftmargin=14pt}

\sethlcolor{lightgray}

\graphicspath{{figures/}}

\newcommand{\mypara}[1]{%
\vspace{2mm} \noindent
\textbf{#1.} \;
}
\newcommand{\smallpara}[1]{%
\vspace{1mm}
\emph{#1}.\;
}

\definecolor{StrongBlue}{HTML}{1F78B4}
\definecolor{BrightRed}{HTML}{E31A1C}
\definecolor{MediumGreen}{HTML}{33A02C}
\definecolor{WarmOrange}{HTML}{FF7F00}
\definecolor{DeepPurple}{HTML}{6A3D9A}

\definecolor{BlueColor}{HTML}{1F78B4}
\definecolor{RedColor}{HTML}{E31A1C}
\definecolor{OrangeColor}{HTML}{FF7F00}
\definecolor{YellowColor}{HTML}{FFC107}

\definecolor{colProc}{named}{RedColor} %
\definecolor{colRest}{named}{MediumGreen} %
\definecolor{colVM}{named}{StrongBlue}   %

\definecolor{etcd}{named}{StrongBlue}
\definecolor{AirIndex}{named}{BrightRed}
\definecolor{slsmr}{named}{WarmOrange}

\newcommand{\api}[1]{\texttt{#1}\xspace}

\newcommand{\system}{AirDB\xspace}
\newcommand{\ssmr}{S$^2$MR\xspace}

\newcommand*{\belowrulesepcolor}[1]{%
  \noalign{%
    \kern-\belowrulesep
    \begingroup
    \color{#1}%
    \hrule height\belowrulesep
    \endgroup
  }%
}
\newcommand*{\aboverulesepcolor}[1]{%
  \noalign{%
    \begingroup
    \color{#1}%
    \hrule height\aboverulesep
    \endgroup
    \kern-\aboverulesep
  }%
}

\definecolor{comment-color}{rgb}{0.5,0.1,0.1}
\let\oldnl\nl
\newcommand{\nonl}{\renewcommand{\nl}{\let\nl\oldnl}}

\definecolor{GreenColor}{HTML}{004D40}
\definecolor{MagentaColor}{HTML}{8E24AA}
\definecolor{vintageblack}{HTML}{484043}
\definecolor{vintageyellow}{HTML}{FFC805}
\definecolor{vintagegreen}{HTML}{38A528}
\definecolor{vintageorange}{HTML}{FF4D25}

\definecolor{airdbColor}{HTML}{e21d23}
\definecolor{airdbMockColor}{HTML}{E66100}
\definecolor{mysqlColor}{HTML}{004D40}
\definecolor{postgresColor}{HTML}{1E88E5}
\definecolor{azuresqlColor}{HTML}{FFC107}

\definecolor{DP}{named}{DeepPurple}
\definecolor{MG}{named}{MediumGreen}

\begin{document}
\pagestyle{plain}
\date{}

\title{\Large \bf Eliminating Servers in State Machine Replication}

\author{Dohyun Park}
\email{dohyunp2@illinois.edu}
\affiliation{%
  \institution{UIUC}
  \city{}
  \country{}
}

\author{Wenjie Hu}
\email{wenjiehuhippo@cs.wisc.edu}
\affiliation{%
  \institution{University of Wisconsin--Madison}
  \city{}
  \country{}
}


\author{Mahesh Balakrishnan}
\authornote{Work done prior to author's affiliation with Meta}
\email{maheshb@gmail.com}
\affiliation{%
  \institution{Meta}
  \city{}
  \country{}
}

\author{Yongjoo Park}
\email{yongjoo@illinois.edu}
\affiliation{%
  \institution{UIUC}
  \city{}
  \country{}
}

\begin{abstract}
	State Machine Replication (SMR) underpins many stateful systems, including distributed databases, coordination services, and metadata stores.
Fundamentally, SMR requires state to be materialized from a durable total order of commands.
Existing SMR systems rely on servers to materialize, locally store, and serve snapshots of state; while the durable total order either resides on the same servers, or on an external, disaggregated shared log.
Unfortunately, servers act as brittle, expensive choke-points in such systems, causing service unavailability when they fail; incurring provisioning latency on cold starts; and red-lining when exposed to bad clients.
In this paper, we propose Serverless SMR (\ssmr{}), where both the materialized state and the durable total order are stored on disaggregated storage; clients concurrently maintain a \emph{single copy} of materialized state on external storage.
Our key technical insight is that the shared log abstraction is useful not just for consistently ordering commands in SMR, but also for enabling distributed materialization, thus becoming the only interface for achieving SMR.
We demonstrate \ssmr{} by developing \system{}, a transactional key-value store.
In our evaluation, \system{} over \ssmr{} achieves higher availability than existing SMR-based systems like etcd over Raft: it is oblivious to server failures and bad client activity; and spins up 10$\times$-45$\times$ faster from a cold start.

\end{abstract}

\maketitle

\section{Introduction}
\label{sec:intro}

State Machine Replication (SMR)~\cite{lamport_time_1978,schneider_implementing_1990} is a foundational building block for stateful services, enabling durability and availability for production systems such as distributed databases and key-value stores (e.g., Spanner~\cite{corbett2013spanner}, ZippyDB~\cite{zippydb}, CockroachDB~\cite{taft2020cockroachdb}, YugabyteDB~\cite{yugabytedb}, Oceanbase~\cite{oceanbase},
	    TiDB~\cite{tidb}), as well as
	    coordination services (e.g., Chubby~\cite{burrowschubby}, etcd~\cite{etcd}, ZooKeeper~\cite{hunt_zookeeper_2010}, Delos~\cite{balakrishnan2020virtual}, KRaft~\cite{kraft}).
In SMR, arbitrary state is replicated with strong consistency by generating and materializing a durable total order (or equivalently, a \textit{shared log}) of commands.
SMR is exceptionally versatile: the replicated state can consist of any black-box deterministic state machine (e.g., single-node databases, hierarchical namespaces, block stores, etc.), while the shared log can be stored via any consensus protocol: for example, an SMR system might replicate MySQL database instances using Raft (e.g., Meta's MyRaft~\cite{myraft}).
In addition, SMR systems can support either single-master or multi-master operation; while the shared log can store either inputs / logical commands or outputs / physical change records.

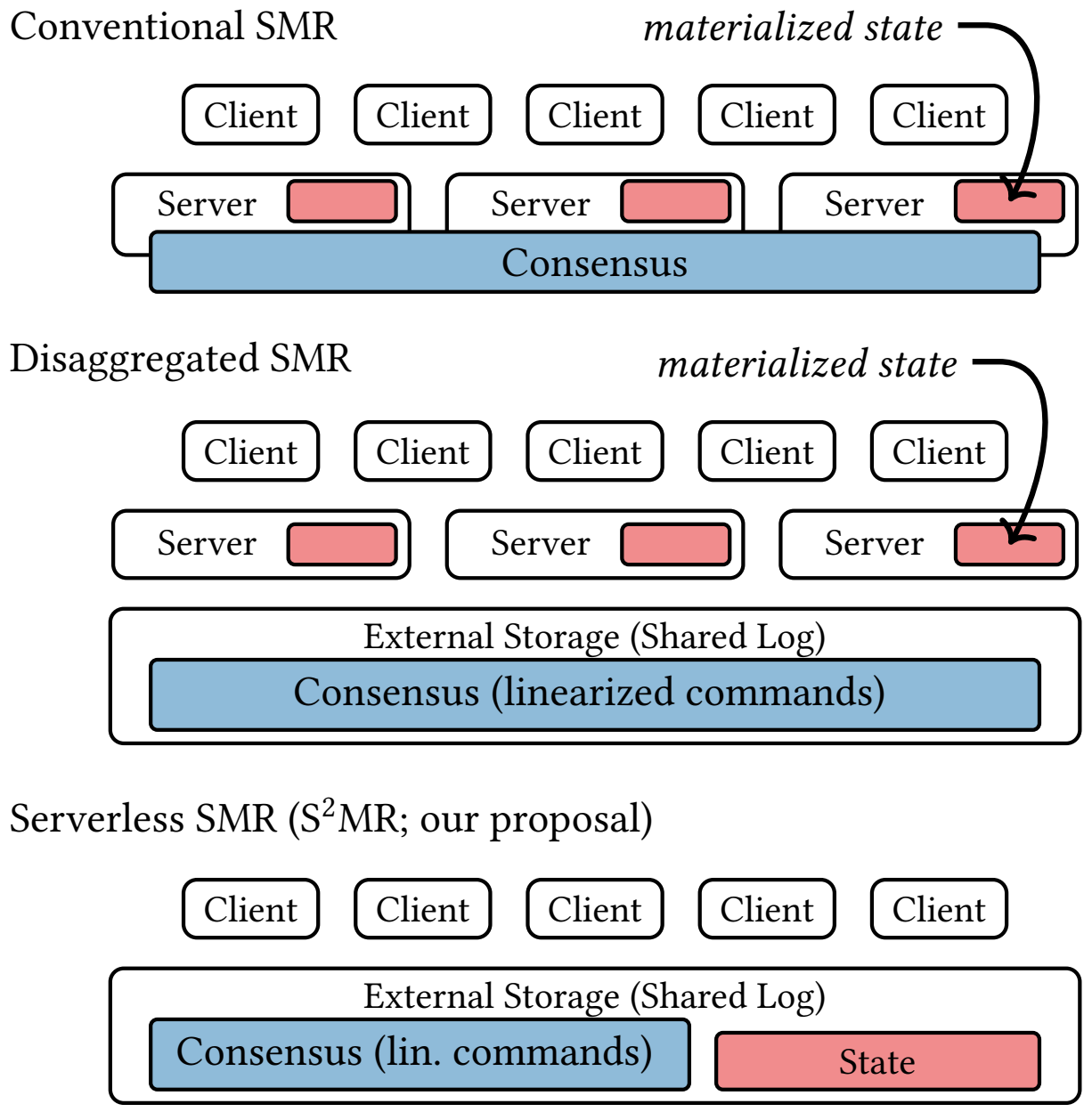
\begin{figure}[t]

\tikzset{
base/.style={
    draw=black,rounded corners=1mm,minimum width=10mm,thick,
    font=\footnotesize,align=center,
    minimum height=4mm,
},
mylabel/.style={
    font=\small
},
myserver/.style={
    base,font=\footnotesize,minimum width=22mm,minimum height=6mm,
    label={[yshift=-4.5mm,xshift=-4mm,font=\footnotesize]above:{Server}}
},
consensus/.style={
    base,fill=white,minimum width=66mm,minimum height=3mm,rounded corners=0.5mm,
    fill=StrongBlue!50!white,font=\small
},
mystate/.style={
    base,minimum height=3mm,minimum width=8mm,rounded corners=0.5mm,
    fill=BrightRed!50!white
},
myserver2/.style={
    base,font=\small,minimum width=22mm,minimum height=5mm,
    label={[yshift=-4.7mm,xshift=-4mm,font=\footnotesize]above:{Server}}
},
sharedlog/.style={
    base, minimum width=72mm,minimum height=10mm,
    label={[yshift=-5mm,font=\footnotesize]above:{External Storage (Shared Log)}}
}
}

\centering
\begin{tikzpicture}

\def\xg{0.25}
\node[base] (C1) {Client};
\node[base,anchor=west] (C2) at ($(C1.east)+(\xg,0)$) {Client};
\node[base,anchor=west] (C3) at ($(C2.east)+(\xg,0)$) {Client};
\node[base,anchor=west] (C4) at ($(C3.east)+(\xg,0)$) {Client};
\node[base,anchor=west] (C5) at ($(C4.east)+(\xg,0)$) {Client};

\node[myserver,anchor=north] (S2) at ($(C3.south)+(0,-0.2)$) {};
\node[myserver,anchor=east] (S1) at ($(S2.west)+(-\xg,0)$) {};
\node[myserver,anchor=west] (S3) at ($(S2.east)+(\xg,0)$) {};

\node[consensus,anchor=north] (CS) at ($(S2.south)+(0,0.2)$) {Consensus};

\node[mystate,anchor=east] (M1) at ($(S1.east)+(-0.1,0.1)$) {};
\node[mystate,anchor=east] (M2) at ($(S2.east)+(-0.1,0.1)$) {};
\node[mystate,anchor=east] (M3) at ($(S3.east)+(-0.1,0.1)$) {};

\node[mylabel,anchor=south west] (L1) at ($(C1.north west)+(-1.4,0.2)$) {Conventional SMR};

\node[font=\small\itshape,anchor=west] (MS) at ($(L1.east)+(2,0)$) {materialized state};
\draw[->,very thick] (MS.east) -- ++(0.2,0) to[out=0,in=60] (M3.center);

\def\yg{2.5}
\node[mylabel,anchor=west] (L2) at ($(L1.west)+(0,-\yg)$) {Disaggregated SMR};

\node[base] (C1) at ($(C1)+(0,-\yg)$) {Client};
\node[base,anchor=west] (C2) at ($(C1.east)+(\xg,0)$) {Client};
\node[base,anchor=west] (C3) at ($(C2.east)+(\xg,0)$) {Client};
\node[base,anchor=west] (C4) at ($(C3.east)+(\xg,0)$) {Client};
\node[base,anchor=west] (C5) at ($(C4.east)+(\xg,0)$) {Client};

\node[myserver2,anchor=north] (S2) at ($(C3.south)+(0,-0.2)$) {};
\node[myserver2,anchor=east] (S1) at ($(S2.west)+(-\xg,0)$) {};
\node[myserver2,anchor=west] (S3) at ($(S2.east)+(\xg,0)$) {};

\node[sharedlog,anchor=north] (SL) at ($(S2.south)+(0,-0.2)$) {};
\node[consensus,anchor=south] (CS) at ($(SL.south)+(0,0.1)$) {Consensus (linearized commands) };

\node[mystate,anchor=east] (M1) at ($(S1.east)+(-0.1,0)$) {};
\node[mystate,anchor=east] (M2) at ($(S2.east)+(-0.1,0)$) {};
\node[mystate,anchor=east] (M3) at ($(S3.east)+(-0.1,0)$) {};

\node[font=\small\itshape,anchor=west] (MS) at ($(L2.east)+(2,0)$) {materialized state};
\draw[->,very thick] (MS.east) -- ++(0.2,0) to[out=0,in=60] (M3.center);

\def\ygg{3.4}
\node[mylabel,anchor=west] (L3) at ($(L2.west)+(0,-\ygg)$) {Serverless SMR (\ssmr; our proposal)};

\node[base] (C1) at ($(C1)+(0,-\ygg)$) {Client};
\node[base,anchor=west] (C2) at ($(C1.east)+(\xg,0)$) {Client};
\node[base,anchor=west] (C3) at ($(C2.east)+(\xg,0)$) {Client};
\node[base,anchor=west] (C4) at ($(C3.east)+(\xg,0)$) {Client};
\node[base,anchor=west] (C5) at ($(C4.east)+(\xg,0)$) {Client};

\node[sharedlog,anchor=north] (SL) at ($(C3.south)+(0,-0.2)$) {};
\node[consensus,minimum width=40mm,anchor=south west] (CS) at ($(SL.south west)+(0.3,0.1)$) {Consensus (lin.~commands) };
\node[mystate,minimum width=24mm,anchor=south east] (CS) at ($(SL.south east)+(-0.3,0.1)$) {State};

\end{tikzpicture}

\caption{
Architectural comparison.
Both conventional and disaggregated SMR materialize state on servers.
Under \ssmr, distributed clients materialize state on external storage via a shared log,
    achieving higher availability and competitive latency-throughput at lower infrastructure costs.
}

\label{fig:intro}
\end{figure}

Unfortunately, durable and highly available services based on SMR are notoriously difficult and expensive to build, deploy, and operate~\cite{chandra2007paxos, balakrishnan2024taming}.
Consider the example of a MySQL database replicated via Raft~\cite{myraft}, which requires the replicated state (i.e., MySQL instances) and the shared log (i.e., Raft participants) to be collocated on a single set of servers.
Such a service is typically deployed with five servers storing full copies of state: Raft requires at least three servers to tolerate a single server failure; and five servers to tolerate a correlated failure during a planned server upgrade.
Availability can be compromised by server crashes (e.g., the Raft leader), I/O contention between Raft and MySQL, software bugs in either layer that crash the entire process, etc.
Durability is even more challenging to ensure: a correlated failure (e.g., due to a bad software upgrade) that takes down a majority of nodes can result in data loss (particularly if the underlying scheduler does not guarantee disk durability across node reboots).

\begin{figure}

\begin{tikzpicture}

\def\x{0.8\linewidth}
\def\y{0.4\linewidth}

\node[draw=black,very thick,minimum width=\x,minimum height=\y] (Q) at (0,0) {};

\node[anchor=north,font=\bfseries] at ($(Q.south)+(0,-0.5)$) {Log Placement};
\node[anchor=north] at ($(Q.south)+(-0.25*\x,0.0)$) {server(s)};
\node[anchor=north] at ($(Q.south)+(0.25*\x,0)$) {external/cloud storage};

\node[anchor=south,font=\bfseries,align=left,rotate=90] at ($(Q.west)+(-0.9,0)$) {State Placement};
\node[anchor=south,rotate=90,align=center] at ($(Q.west)+(0,0.25*\y)$) {external/cloud\\[-0.25em] storage};
\node[anchor=south,rotate=90] at ($(Q.west)+(0,-0.25*\y)$) {server(s)};

\draw[semithick,draw=black] (Q.north) -- (Q.south);
\draw[semithick,draw=black] (Q.west) -- (Q.east);

\node[font=\bfseries\small] at ($(Q.center)+(0.25*\x,0.25*\y)$) {\ssmr (ours)};

\node[font=\footnotesize] at ($(Q.center)+(-0.25*\x,0.25*\y)$) {N/A};
\node[font=\footnotesize,text width=0.45*\x] at ($(Q.center)+(-0.25*\x,-0.25*\y)$) {PostgreSQL~\cite{postgresql}, etcd~\cite{etcd}, AzureSQL~\cite{azsql}, Spanner~\cite{corbett2013spanner}, ZooKeeper~\cite{hunt_zookeeper_2010}, CockroachDB~\cite{taft2020cockroachdb}};

\node[font=\footnotesize,text width=0.45*\x] at ($(Q.center)+(0.25*\x,-0.25*\y)$) {LazyKV~\cite{lazylog}, Delos~\cite{balakrishnan2020virtual}, CorfuDB~\cite{corfudb}, Tango~\cite{balakrishnan2013tango}, Scalog~\cite{ding_scalog_2020}};

\end{tikzpicture}

\vspace{-2mm}
\caption{Taxonomy of SMR. We propose \underline{Serverless SMR} (\ssmr), which places both the replication log and materialized state in external storage, eliminating servers.}
\end{figure}

To mitigate these challenges, recent SMR systems have \textit{disaggregated the shared log}, delegating durability entirely to an external (potentially cloud-based) storage service via an explicit shared log API~\cite{bernstein2011hyder,balakrishnan2020virtual, balakrishnan2013tango, ding_scalog_2020, lazylog}.
In such systems (which we call disaggregated SMR), servers do not implement consensus protocols to maintain a shared log; instead, they store the shared log on external storage and locally materialize soft-state snapshots of application state (e.g., a MySQL database).
Servers are no longer responsible for durability: service state can be reconstructed even if all servers fail permanently, as long as the shared log on the external storage service continues to exist.
The key driver is the exceptional durability and availability of shared-log architectures~\cite{balakrishnan2012corfu, lazylog}, which separate log ordering from storage through a dedicated sequencing layer rather than peer-to-peer consensus among materializers.

However, even if disaggregated SMR improves durability, availability remains challenging to ensure.
The availability of the service continues to depend on a server-based materializer layer that can be difficult to harden, for three reasons:
\begin{itemize}
\item{\textbf{Server Failures / Upgrades:} The materialization layer consists of multiple failure-prone components (server software, server hardware, deployment infrastructure such as Kubernetes, etc.). 
Upgrades to service logic (e.g., to support new features) can be frequent and risky~\cite{upgradefailures}.}
\item{\textbf{Slow Start:} The service must trade off cost against availability. Keeping full-state servers ready incurs energy and cloud costs, while starting one on demand requires downloading a snapshot and may violate the latency SLA.}
\item{\textbf{Lack of Server Isolation:} Servers have to ensure their own availability in a multi-tenant environment via complex mechanisms for rate-limiting and quarantining (e.g., to prevent one client's long-running full-DB scan from starving another client's time-critical single-item reads).}
\end{itemize}
In effect, continued reliance on materialization servers for availability makes services complex, brittle, and expensive.

\mypara{Our Proposal} In this paper, we propose eliminating servers entirely: we disaggregate both materialized states and ordered commands onto the shared log.
The key insight is that a shared log's \api{append} supports both distributed command ordering and distributed materialization.
Specifically, \ssmr (\textbf{S}erverless \textbf{S}tate \textbf{M}achine \textbf{R}eplication) stores all state---including the state machine and the command log---entirely in external storage, where the storage can be one of the existing shared-log systems or cloud storage exposing similar APIs (e.g., Azure Blob).
\ssmr inherits the foundational mechanism of the shared log: multiple clients consistently can manage (read/write) the shared state by directly accessing the log.
\ssmr offers an additional benefit by allowing clients to query the state more quickly by inspecting the materialized snapshot without replaying the entire log from scratch.
The materialized snapshots (with collocated indexes) are created by clients in a distributed way.
Notably, only a single logical copy of materialized state is maintained by all clients (as opposed to conventional SMR or disaggregated SMR, where each server maintains its own independent copy).

\ssmr's serverless design offers ``always-on'' availability:
\begin{itemize}
\item{\textbf{No Servers (and hence, no server failures or upgrades): } Without materialization servers, service availability depends only on the underlying shared log.}
\item{\textbf{Fast Start: } The service is immediately responsive and available when a workload initially starts and ramps up, without having to over-provision servers in advance or spin them up on demand.}
\item{\textbf{Better Isolation: } Misbehaving or crashing clients have no availability impact on other clients and little impact on external storage, which is designed to provide isolation for multiple tenants via its simple, low-level storage API.}
\end{itemize}
\ssmr{} offers other practical benefits. It is cheaper for service operators, who no longer operate materialization servers and can push functionality to end clients.
It also requires virtually no deployment effort: a new instance is initialized by creating a separate workspace (e.g., folder) in external storage and distributing access credentials to clients.

The resulting protocol retains the simplicity and strong safety guarantees of shared-log systems: the log defines a linearizable command order, and the materialized state is a deterministic function of that log.
When one client advances the shared materialized view, it reduces the replay work required by all clients.

\looseness=-1 We evaluate \ssmr by building \system{}, a transactional key-value store that provides ACID guarantees with Strict Serializability and Snapshot Isolation~\cite{berenson1995critique}.
AirDB offers the ``always-on'' availability of \ssmr{} along with the additional benefits of low cost and zero deployment effort.
For bursty, infrequently accessed state, it also provides strong semantics and lower latency than conventional designs at the same cost, improving availability under a fixed latency target.

\ssmr{} does come with trade-offs. First, migrating the data layout in external storage requires careful coordination. 
Since \ssmr disaggregates all state onto the shared log while logic resides on the client side, upgrading the data layout can leave clients with an outdated library unable to read the new format. 
To address this, a version number can be encoded as part of the materialized state; upon a major version change, a client must either update its library or replay the entire shared log from scratch, incurring longer cold-start latency. 
Second, \ssmr{} adds round trips between clients and external storage, so it best suits deployments that colocate the two within a data center, such as web servers using a storage service in the same facility.

We make the following contributions in this paper:
\begin{itemize}
    \item{We propose \ssmr{}, a novel SMR variant that eliminates servers and materializes states on disaggregated storage.}
    \item{We describe the design and implementation of \system{} using \ssmr{}, showing that it provides strong semantics.}
    \item{Our evaluation shows that \ssmr{} achieves higher availability than both conventional and disaggregated SMR: it is oblivious to materializer failures and denial-of-service (DoS) attacks on clients; and recovers a 1\,GB state from a cold start 45$\times$ faster than conventional SMR and 10$\times$ faster than disaggregated SMR.}
\end{itemize}

\section{The Evolution of State Machine Replication}
\label{sec:motivation}

State Machine Replication was first described~\cite{lamport_time_1978} by Lamport as a way to replicate any arbitrary state machine using a total order of commands, each of which could be decided, stored, and accessed via an instance of fault-tolerant consensus.
Early descriptions of SMR (e.g., in Lamport's description of Paxos~\cite{lamport_paxos_2001}) typically distinguished between the multiple roles played by potentially different machines: \textit{acceptors} that decide and store the total order; \textit{proposers} that suggest candidate commands to be stored and ordered; and \textit{learners} that play the total order and materialize the state machine.
This explicit separation of roles was further reinforced by foundational theoretical results that mapped the problem of fault-tolerant consensus to the implementation of a durable, strongly consistent write-once register~\cite{fischer_flp_1985}: in this view, SMR involved proposers writing to (and learners reading from) an address space of registers, each of which was stored via a quorum protocol like Paxos on a set of acceptors.

However, this clean separation of roles in SMR began to blur with the deployment of SMR systems in industry such as Chubby (2001)~\cite{burrowschubby}, ZooKeeper (2008)~\cite{hunt_zookeeper_2010}, and Raft (2012)~\cite{raft}. 
Each of these systems combines all three roles into a single \textbf{converged SMR} architecture, typically for performance and ease of small-scale deployment.
In these systems, a single process on each machine acts as learner, proposer, and acceptor; accordingly, every process is responsible for both the durability and availability of the service.

Starting in the 2010s, shared log systems in research~\cite{balakrishnan2012corfu, balakrishnan2013tango, lockerman2018fuzzylog, bernstein2011hyder, ding_scalog_2020, nawab2015chariots, lazylog, speclog} and industry~\cite{balakrishnan2020virtual, balakrishnan2021log, corfudb, wei2017vcorfu, logdevice, verbitski2017amazon, kreps_kafka_nodate, memorydb} reintroduced the clean logical (and physical) separation of roles into SMR. 
Such systems adopt a \textbf{disaggregated SMR} design where a tier of materializing servers uses an explicit shared log abstraction to store and order commands.
In this setup, materialization encompasses the proposer and learner roles; while the shared log API hides the acceptor role.
These systems offer the benefits of layered design: each layer can be scaled independently; optimized and sized separately without contending for shared hardware; and upgraded separately.
In particular, the shared log can be implemented over inexpensive cloud storage.

Splitting SMR in this manner (materializers running on servers + a shared log on external \emph{passive} storage) also draws out stark differences in the nature of the two layers.
First, SMR materializers are \textit{diverse}~\cite{balakrishnan2021log, balakrishnan2013tango}, encompassing key-value stores, SQL databases, queues, hierarchical namespaces, and object stores; while external storage has relatively little diversity.
Second, SMR materializers can be \textit{fast-changing}~\cite{upgradefailures}, with new code rolled out continuously to support new user-facing features, improve performance, and so on; whereas storage is slow-moving.
Third, SMR APIs can be \textit{ill-conditioned}~\cite{balakrishnan2024taming}, allowing a single request to trigger large, arbitrary amounts of work (e.g., full-DB scans); whereas storage offers simple APIs that impose constant work per request (e.g., a read on an address).

\looseness=-1 Given this characterization, it follows that disaggregating SMR provides higher durability than converged SMR: failures in the materializer layer are frequent, but no longer impact durability (only availability); failures in storage that do impact durability are far less likely, because its code and APIs are simpler and slower-changing.

In this paper, we explore the next logical step: can we improve availability by eliminating materializers entirely, splitting their functionality across clients and storage?

\subsection{Understanding SMR Availability. Or: Servers considered Harmful!}

We first consider the ways in which materializers can fail and cause SMR downtime; and why eliminating them helps.

\textit{Hardware failures and network outages:} Machines crash due to kernel panics, hardware errors, and power failures; data centers disconnect from each other. Intelligent placement of replicas (e.g., each one in a different region) ensures that these types of failures map well to classical failure assumptions (e.g., no more than a majority fails). 

\looseness=-1 \textit{Schedulers:} In virtualized cloud environments, schedulers such K8s and Twine can add an additional layer of faults~\cite{barletta2024mutiny}, either by preempting physical hosts for maintenance tasks during correct operation, or by interacting in unexpected ways with service logic (e.g., marking healthy hosts as failed; or failing to wait for the service to become healthy during a rolling upgrade). 
Anecdotal reports~\cite{schedulerbadness1, schedulerbadness2, schedulerbadness3} show that stateful services encounter these failure modes with particular frequency and severity.

\looseness=-1 \textit{Software upgrades:} Materializers are constantly upgraded to support new features; each such upgrade is fraught. Zhang et al.~\cite{upgradefailures} provide a comprehensive characterization of upgrade failures in production, highlighting the frequency and severity of such failures.
New code can have bugs that only manifest once a rolling upgrade has completed, resulting in correlated failures (e.g., hangs or crash-loops) of all replicas.
A single node failure during a rolling upgrade can cause unavailability for a three-replica deployment; or even data loss (if the host is swapped out during the upgrade and the deployment uses local disks).

\textit{Bad tenants:} Multi-tenancy is difficult to achieve for diverse services with complex APIs. 
Materialization servers can be overwhelmed by rogue tenants that impose large amounts of work: grabbing locks, multi-versioned state management, scrambling caches, and stressing metadata structures.

\looseness=-1 The common theme is that clients must depend on a layer of servers that are stateful and hard to manage, upgrade, and protect. Ideally, a client accessing state would depend only on its own fate and highly reliable low-level services such as networking and external storage, including AWS S3 and Azure Blob Storage.

\begin{figure*}

\tikzset{
base/.style={
    draw=black,rounded corners=1mm,thick,
    font=\footnotesize,align=center,
    minimum width=10mm, minimum height=6mm,
},
mycommand/.style={
    font=\scriptsize, inner sep=0mm
}
}

\begin{tikzpicture}

\def\xg{0.2}
\def\lshift{3.2}

\node[base,opacity=0] (C1) {Client};
\node[base,anchor=west] (C2) at ($(C1.east)+(\xg+1.0,0)$) {Client};
\node[base,anchor=west] (C3) at ($(C2.east)+(\xg,0)$) {Client};
\node[base,anchor=west,draw=none,font=\large\bfseries,minimum width=5mm] (C4) at ($(C3.east)+(\xg,0)$) {. . .};
\node[base,anchor=west,minimum width=60mm,font=\footnotesize] (C5) at ($(C4.east)+(\xg,0)$) {};

\node[anchor=west,font=\scriptsize,fill=MediumGreen,fill opacity=0.3,text opacity=1,inner sep=0.5mm,draw=MediumGreen] 
    (index1) at ($(C5.west)+(0.1,0)$) {$\mathcal{I} \left( \phi_{(0,7]} \right)$};

\node[anchor=west,font=\scriptsize,fill=StrongBlue,fill opacity=0.3,text opacity=1,inner sep=0.5mm,draw=StrongBlue] 
    (index2) at ($(index1.east)+(0.1,0)$) {$\mathcal{I} \left( \phi_{(7,14]} \right)$};

\node[anchor=west,font=\scriptsize,fill=DeepPurple,fill opacity=0.3,text opacity=1,inner sep=0.5mm,draw=DeepPurple] 
    (index3) at ($(index2.east)+(0.1,0)$) {$\mathcal{I} \left( \phi_{(14,21]} \right)$};

\node[anchor=west,font=\scriptsize,fill=white,fill opacity=0.3,text opacity=1,inner sep=0.5mm,draw=black] 
    (index4) at ($(index3.east)+(0.1,0)$) {$\left( \phi_{(21,24]} \right)$};

\node[anchor=north east,font=\small\itshape,inner sep=0mm] (L1) 
    at ($(C2.north west)+(-0.4,0)$) {Distributed clients};

\node[base,anchor=north west,minimum width=120mm, minimum height=4mm,rounded corners=0mm] (CL) at ($(C1.south west)+(0,-0.5)$) {};
\foreach \x in {0.5, 1.0, ..., 11.5} {
  \draw[thick] ($(CL.north west)+(\x,0)$) -- ($(CL.south west)+(\x,0)$);
}

\foreach \i in {1,...,21} {

  \pgfmathsetmacro{\xl}{0.5*(\i-1)}
  \pgfmathsetmacro{\xr}{0.5*\i}

  \ifnum\i<8
    \def\cellcolor{MediumGreen}
  \else\ifnum\i<15
    \def\cellcolor{StrongBlue}
  \else
    \def\cellcolor{DeepPurple}
  \fi\fi

  \fill[\cellcolor, fill opacity=0.3]
    ($(CL.north west)+(\xl, 0)$)
    rectangle
    ($(CL.south west)+(\xr, 0)$);
}

\def\special{1,2,7,8,9,14,15,16,21,22,23,24}

\foreach \i in {1,...,24} {
  \node[mycommand] at ($(CL.north west)+(0.5*\i - 0.25, -2mm)$)
    {
      \ifnum\pdfmatch{,\i,}{,\special,}=1
        $\sigma_{\i}$%
      \else
        $\cdots$%
      \fi
    };
}

\node[anchor=west] (L2) at ($(CL.west)+(-\lshift,0)$) {The command log};

\node[base,anchor=north west,minimum width=60mm, minimum height=4mm,
      rounded corners=0mm] (SL) at ($(CL.south west)+(0,-0.3)$) {};

\fill[MediumGreen, fill opacity=0.3]
  (SL.north west)
  rectangle
  ($(SL.south west)+(20mm,0)$);

\fill[StrongBlue, fill opacity=0.3]
  ($(SL.north west)+(20mm,0)$)
  rectangle
  ($(SL.south west)+(40mm,0)$);

\fill[DeepPurple, fill opacity=0.3]
  ($(SL.north west)+(40mm,0)$)
  rectangle
  (SL.south east);

\foreach \x in {20mm, 40mm} {
  \draw[thick] ($(SL.north west)+(\x,0)$) -- ($(SL.south west)+(\x,0)$);
}

\node[anchor=west] (L3) at ($(SL.west)+(-\lshift,0)$) {The state log};

\node[mycommand,font=\scriptsize]
  at ($(SL.north west)+(10mm, -2mm)$) {$\phi_{(0,7]}$};

\node[mycommand,font=\scriptsize]
  at ($(SL.north west)+(30mm, -2mm)$) {$\phi_{(7,14]}$};

\node[mycommand,font=\scriptsize]
  at ($(SL.north west)+(50mm, -2mm)$) {$\phi_{(14,21]}$};

\node[minimum width=18mm] (PadRight) at ($(CL.east)+(10mm,0)$) {};

\node[
  draw=gray,
  draw opacity=0.5,
  ultra thick,
  rounded corners=1mm,
  inner sep=3pt,
  fit=(L2) (L3) (CL) (SL) (PadRight)
] (SharedBox) {};

\node[anchor=south west, font=\small\bfseries, align=center, inner sep=0mm, text=gray] 
    at (SharedBox.north west) {The Shared Log / Cloud Storage};

\node[font=\small,draw=black,thick,anchor=west,opacity=0] (newcommand) at ($(C5.east)+(0.4,-0.2)$) {$\sigma$};
\draw[->,very thick]
  (C5.east)
  to[bend left=30] (newcommand.center)
  to[bend left=30] (CL.east);
\node[font=\small,draw=black,thick,fill=white] at (newcommand) {$\sigma$};

\node[font=\footnotesize\itshape, anchor=south west]
  at ($(newcommand.north east)+(-0.1,0.0)$) {new command};

\coordinate (StateLastEast) at ($(SL.east)+(0, 0)$);

\node[
  base,
  anchor=west,
  minimum width=25mm,
  minimum height=6mm,
  rounded corners=0.8mm,
  fill=DeepPurple,
  fill opacity=0.15,
  text opacity=1,
  draw=DeepPurple,
  minimum width=45mm,
  font=\footnotesize
] (SLzoom) at ($(SL.east)+(12mm,0)$) {compaction of $\langle \sigma_{15}, \ldots, \sigma_{21} \rangle$, \; and index $\mathcal{I}\left( \phi_{(14,21]} \right)$};

\draw[draw=none,fill=gray,fill opacity=0.5] (StateLastEast) to[out=30,in=180] ++(0.9,0.15) -- ++(-0.1,0.15) -- ($(SLzoom.west)$)
    -- ++(-0.35,-0.3) -- ++(0.1,0.2) to[out=160,in=20] ($(StateLastEast)+(0,-0.05)$);
\node[text=gray,font=\scriptsize,anchor=west] at ($(StateLastEast)+(0.1,0.3)$) {zoom-in};

\end{tikzpicture}

\newcommand{\Ibox}{%
  \tikz[baseline=-0.5ex]{%
    \node[
      draw=black,
      fill=MediumGreen,
      fill opacity=0.3,
      text opacity=1,
      font=\scriptsize,
      inner sep=0.2mm
    ] {$\mathcal{I}\left( \phi_{(0,7]} \right)$};%
  }%
}

\newcommand{\Sbox}{%
  \tikz[baseline=-0.5ex]{%
    \node[
      draw=black,
      font=\scriptsize,
      inner sep=0.4mm
    ] {$\phi_{(21,34]}$};%
  }%
}

\caption{\ssmr data organization. Authenticated clients store commands and states in the shared log. 
To read state, clients access materialized snapshots using indexes 
(e.g., \protect\Ibox)
with a partial state (e.g., \protect\Sbox) derived from non-materialized commands. 
To update the state, clients append new entries (e.g., $\sigma$) to the command log. Conditional commands (e.g., ``withdraw if balance is sufficient'') are validated by checking the command log to ensure the condition held at the time the command was logged.
Snapshots are created in a timely and incremental manner by distributed clients.
}
\label{fig:ssmr:data}
\end{figure*}
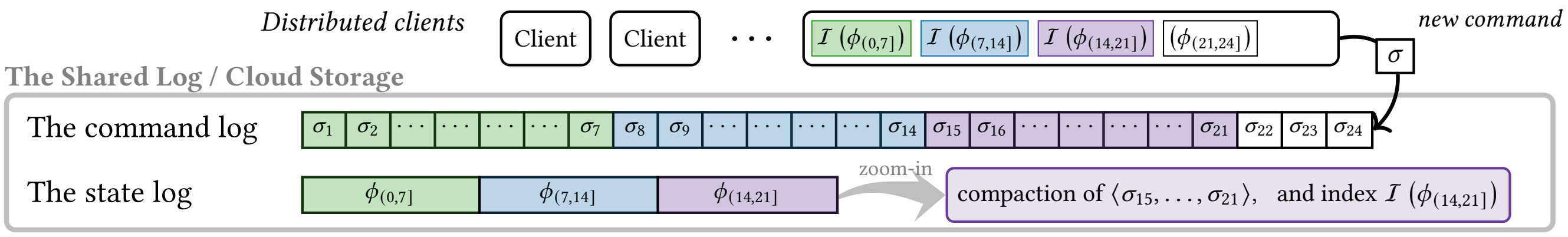

\textbf{Why not just run unmodified SMR on clients?} At first glance, the solution appears simple: eliminate servers and move functionality to clients.
In the context of disaggregated SMR, each client could act as a materializer over the shared log.
Such a solution is trivially correct: the shared log continues to be the source of consistency and durability.
It also definitively eliminates server failures as a source of unavailability -- if a client fails, it only impacts itself -- and provides protection against the activity of other clients.
However, it does not provide the ``fast start'' property: conventional materializers are required to download full snapshots (which can run into multiple GB) before they become operational; if clients are ephemeral, the time-to-first-operation can be prohibitive.
In addition, clients may often not have access to local durable storage, turning simple reboots or preemptions into cold starts.
Finally, requiring each client to redundantly materialize the same shared log (and store snapshots periodically in remote storage) can be onerous, especially if clients are critical services in their own right.
In the other direction, nominating a single client as a designated materializer can be efficient but introduce a new source of unavailability.
Overall, running server-side SMR logic naively on clients might make services \textit{less} available rather than more.

Instead, we need a solution that can provide all three availability properties -- no servers, client isolation, and fast start -- without constraining or overloading clients.
Specifically, we need a new type of client-driven SMR that:
\begin{itemize}
\item{avoids durable local storage on clients.}
\item{reduces overall materialization and snapshot storage load.}
\item{distributes materialization load across clients.}
\end{itemize}

\section{Abstraction: Serverless SMR}

We now describe the \ssmr{} abstraction. \ssmr concerns the class of \emph{state} such that its subset (i.e., partial state) can be efficiently accessed through indexes (e.g., key-value list, inverted index, tree, graph).
The \ssmr architecture involves three parties: the client-side library, \emph{the command log} (i.e., an instance of the shared log), and \emph{the state log} (i.e., another instance of the shared log).
The role of the command log is identical to the conventional role of the shared log: it linearizes commands and serves as the groundtruth state.
\ssmr's difference lies in the way that the client-side library manages snapshots using the state log (see \cref{fig:ssmr:data}).

\textbf{The client-side library} (or \emph{the library}) serves the application clients (e.g., \texttt{get/put} of a key-value store) by communicating with the command log and the state log.
Clients may read the current state or issue a new command to mutate the state; all are coordinated through the library.
Clients may run on multiple machines, all sharing the same global state; the library runs on the same machine as each client. 
From the application's perspective, the library looks like an embedded database such as SQLite~\cite{sqlite} or RocksDB~\cite{rocksdb}, but acts like a distributed database such as etcd~\cite{etcd} or ZooKeeper~\cite{hunt_zookeeper_2010}.

\textbf{The command log} stores linearized commands, which are originally issued by the client and are appended to the command log via the library.
Each command may mutate the state by directly setting values (e.g., \texttt{mykey=val}) or by issuing an operation with optional conditions (e.g., \texttt{balance-=100 if balance>=100}).
Processing a conditional operation requires two round-trips to ensure no concurrent commands have invalidated the condition.
Our system supports both cases. Our evaluation includes a multiple-round-trip scenario, i.e., transaction processing based on snapshot isolation~\cite{berenson1995critique}.

\textbf{The state log} stores incremental snapshots, which are created and appended by the library.
The snapshots allow fast partial state retrieval (e.g., using indexes).
The size of the indexes (e.g., storing key-offset pairs) may be significantly smaller than the original command log.
Each incremental snapshot is created by playing a certain number of commands.
The new snapshot is then added to the state log, using the \emph{timely snapshot} mechanism described in \cref{sec:ssmr:compaction}.
This snapshot creation consumes clients' compute resources, just like RocksDB and SQLite do, while in \ssmr, the resource usage is distributed across multiple clients.

Circling back to \textbf{the library}, retrieving and mutating the current state requires data retrieval from two sources in general: (1) partial states from snapshots and (2) the new commands that have been added since the last snapshot creation. 
Fetching these data is efficient because (1) retrieving partial states is performed efficiently with indexes; and (2) the number of new commands is automatically rate-limited if the data ingestion rate is too high (\cref{sec:ssmr:compaction}).

Finally, \ssmr can easily ensure system-wide correctness because it relies on a shared log to linearize commands and manage snapshots. All clients observe the same state by deterministically applying logged commands or reading their materialized results from the snapshots.

\subsection{High-level Workflow}

The library performs three types of operations: (1) appending commands, (2) updating the local state, and (3) creating snapshots.
Each operation issues a bounded number of storage API calls and transfers a bounded number of commands or key-value pairs. Together, these operations implement higher-level application logic, including the database transactions described next.

First, appending a new command to the command log requires a \emph{single} append API call to the command log.

Second, the current state can be constructed by combining (1) materialized snapshots, i.e., the state log, and (2) the commands that have not been materialized yet, i.e., the \emph{tail} of the command log.
Suppose we need to update the value for a key.
Reading it from the state log (i.e., snapshots) leverages indexes, allowing asymptotically logarithmic data accesses.
Reading from the command log is expensive, especially at startup, if there are many non-materialized commands.
\ssmr bounds this by introducing the \emph{timely snapshot} policy; it ensures that all libraries actively create snapshots until the amount of non-materialized commands shrinks enough.

Finally, the library materializes new commands and stores them in the state log as a new incremental snapshot. 
The incremental snapshot only contains the results of applying those new commands. For this reason, the incremental snapshot is small in size, allowing efficient in-memory construction and snapshot transfer to the state log.

\subsection{Data Model}

Let $\mathcal{L} = \langle \sigma_1, \sigma_2, \ldots, \sigma_n \rangle$ be the command log.
The transition function $\rho$ applies a command, constructing a new state $\Phi_{(0,i]} = \rho(\sigma_i; \Phi_{(0,i-1]})$;
$\Phi_{(0,0]}$ is the empty state.
Reading a key-value from $\Phi$ requires only logarithmic I/O with respect to the size of $\Phi$, using a conventional tree-structured index.

\emph{Partial state} $\Phi_{(k,j]}$ satisfies $\Phi_{(0,j]} = \Phi_{(0,k]} \oplus \Phi_{(k,j]}$, where $\oplus$ indicates that the right-operand state \emph{upserts} the left state.
$\Phi_{(k,j]}$ can be constructed efficiently without fully downloading $\Phi_{(k,j]}$: only necessary key-values can be retrieved using indexes while applying $\langle \sigma_{k+1}, \sigma_{k+2}, \ldots, \sigma_j \rangle$.
Specifically, it requires $O\left( (j-k) \cdot \log(|\Phi_{(0,k]}|) \right) $ I/O costs.
The constructed partial states serve as incremental snapshots and are stored in the state log with necessary metadata.
A merged state, e.g., $\phi_{(0,j]}$, is again a \emph{state}, which can then be merged with another state, say $\phi_{(j,l]}$, to form yet another state, i.e., $\phi_{(0,l]}$. This property allows recursive compaction.

\subsection{Distributed Materialization}
\label{sec:ssmr:compaction}

Minimizing the number of non-materialized commands ensures low start-up cost.
To reduce them, snapshots should be created promptly and reliably.

\mypara{Policy}
The library initiates snapshot creation (i.e., \emph{compaction}) if there are more than $B$ non-materialized commands. Specifically, it maintains the \emph{timely snapshot} invariant.

\begin{definition}[Timely Snapshot]
$\mathcal{L} = \langle \sigma_1, \ldots, \sigma_n \rangle$ is the command log.
$\mathcal{S} = \left\langle \Phi_{(0,i_1]}, \ldots, \Phi_{(i_{k-1}, i_k]} \right\rangle$ is the state log.
\emph{Timely snapshot} ensures non-materialized commands cannot grow too much; that is, $n - i_k \le (1 + \epsilon) B$, where $\epsilon$ is a small number to allow a soft limit (e.g., 0.2).
\end{definition}

\vspace{-1mm}
If $n - i_k \ge B$, the library initiates snapshot creation immediately; once $n - i_k$ exceeds $(1 + \epsilon) B$, it stops accepting new commands until enough have been materialized.

\mypara{Coordination}
The library within any distributed clients may participate in snapshot creation.
This may cause \emph{redundancy}; that is, some snapshots may be created and added multiple times.
It is desirable to avoid it.

When a library recognizes the need for snapshot creation, it leverages the state log to ensure no other libraries are in the process of creating the same or overlapping snapshots.
Specifically, a library begins snapshot creation by appending a \emph{begin-snapshot} command to the state log, indicating (1) the range of commands it aims to compact and (2) Time to Live (TTL). Then, the library immediately checks the state log to confirm there are no conflicting \emph{begin-snapshot} commands that appear earlier in the log.
Once confirmed, the library begins actual snapshot creation, which is completed when the declared snapshot is added to the state log.

\mypara{Fault-tolerance}
Compaction may not complete in time, which can be identified.
Suppose a library $L_1$ appended a \emph{begin-snapshot} command and then soon crashed.
Another library $L_2$ finds $L_1$'s \emph{begin-snapshot} command at $L_2$'s local time $T$. If no corresponding snapshot appears until $L_2$'s local time $T + \text{TTL}$, $L_2$ may begin compaction by appending its own \emph{begin-snapshot}.
This doesn't require any global clocks.

\subsection{Small, Bounded I/O}

We analyze the amount of data the library must fetch from the storage during its life cycle. In short, the total I/O cost is (only) linear with respect to the number of reads and writes, and is sublinear with respect to the state size.

A client starts at time $T_1$,
    when $\mathcal{L}$ contains $n_1$ commands.
The client is alive for $d$ seconds, during which $\mathcal{L}$ receives $m_a$ commands (in total from distributed clients) per second on average.
When the client terminates, $\mathcal{L}$ contains $n_e$ commands, where $n_e = n_1 + m_a \cdot d$.

At $T_2 (> T_1)$, the client checks the state log to know that snapshots have been created for the $[\sigma_1, \ldots, \sigma_{n'}]$ commands where $n' \le n_1$.
At $T_3 (> T_2)$, the client read all currently existing commands positioned after $n'$.
$T_3 - T_1$ is the start-up time.
Now, the client is ready to serve client requests.
While serving requests, the client continuously reads new commands, totaling $(n_e - n')$ commands across its lifetime.

The client maintains two kinds of local states: (1) snapshot key set, and (2) the compacted tail.
The snapshot key set is the merge of all the indexes of the snapshots that existed at $T_1$;
    formally, $\mathcal{I}(\Phi_{(0,i_1)}) \oplus \cdots \oplus \mathcal{I}(\Phi_{(i_{k-1},i_k)})$,
        where $\mathcal{I}$ indicates the index (e.g., key-offset pairs) of a snapshot.
The result of the merge is another index; it tells us which snapshot contains the latest value of a given key. 
The compacted tail is a partial state created by applying the commands positioned after $n'$, i.e., the last snapshot point; unlike the index, it also stores the values themselves, not just keys.

If the client makes an average of $m_r$ read requests per second,
its per-second I/O cost is $O(m_r \cdot \log(|\Phi|) + m_a)$,
    where $|\Phi|$ indicates the size of its key set.
That is, the I/O cost is nearly proportional to the number of read/append requests, scaling sublinearly with respect to the state size.

\section{AirDB: \ssmr-based Key-Value Store}

Built on \ssmr, \system provides a transactional key-value store interface.
It runs inside distributed clients, manages globally shared state on cloud storage, and is implemented in Go in roughly 6.6K lines of code.

We describe how \system extends \ssmr abstraction in terms of (1) command and state transition; (2) transaction-aware state construction, and (3) the shared-log storage on top of cloud storage.
Finally, we present optimization details.

\subsection{A Command is A DB Transaction}

\system can be configured at deployment time to support either Snapshot Isolation (SI) or Strict Serializability (SS).
Our transaction protocol implements optimistic concurrency control over the shared log, similar to prior shared log databases~\cite{bernstein2011hyder}.
A client begins a transaction $T$ by synchronizing with the tail of the command log.
It uses this log offset as the snapshot timestamp $s(T)$ for the transaction.
The client then speculatively executes the transaction by issuing reads and buffering writes, generating an \textit{intention} that describes either the read set (for SS) or the write set (for SI).

To commit the transaction, we first append the intention as a command to the command log.
The position of the intention in the log is the commit timestamp $c(T)$ for the transaction.
We then determine if the transaction commits or aborts by inspecting the state---the command log and state log--- at the commit timestamp to see if any conflicting transactions have committed since our snapshot timestamp.
In the case of SI, this requires checking whether any keys in the $T$'s write set have been written between $s(T)$ and $c(T)$.
In the case of SS, this checks whether any keys in $T$'s read set have been written between $s(T)$ and $c(T)$.
Each transaction intention is appended as a single log entry, whose size is bounded by the maximum append size of the underlying shared log (e.g., 4\,MB for Azure Append Blob). Transactions exceeding this limit must be split. 

Checking conflicts---\emph{any updates on my read set (SS) or write set (SI)}---requires historical data. \system's (partial) state representation allows this check in a memory-efficient way, without the full command history, as described below.

\subsection{State Representation}

\system strictly follows the \ssmr data model: (1) applying a new command $\sigma_i$ to the previous state $\Phi_{(0,i-1]}$ constructs a new state $\Phi_{(0,i]}$; and (2) two partial states $\Phi_{(0,k]}$ and $\Phi_{(k,j]}$ can be merged into a single state $\Phi_{(0,j]}$.

\emph{Partial State.}\;
\looseness=-1 The state representation should allow conflict checks. Simply storing the latest key-value pairs is insufficient since we cannot answer whether a key was modified between two offsets. 
However, storing the full change history for every key is memory-inefficient; thus, we should avoid it.

We observe that under \ssmr, a new transaction's commit time is always greater than any of the transactions that have been applied to existing states. 
Leveraging this, \system stores the (single) last modification time for each key. This allows correct conflict checks, as follows.
For any key in $T$'s read-set (for SS) or write-set (for SI), if the last modification timestamp is within $(s(T), \text{now})$, we can say there exists at least one conflicting transaction; thus, $T$ should be aborted. Vice versa, if the last modification time of a key is before $s(T)$, we can say that the key has not been modified since $T$ observed it in its snapshot (SS) or wrote it (SI); thus, $T$ is committed.

\emph{Merge.}\;
Merging two states, $\Phi_{(0,k]}$ and $\Phi_{(k,j]}$, is straightforward. The key-value pairs from $\Phi_{(k,j]}$ overwrite any existing key-value pairs, including the last modification time.
Note that merging doesn't need to involve conflict checks. Each state only contains the results of committed transactions in relation to their respective prior transactions.

In summary, \system requires $O(|\Phi|)$ memory in maintaining (partial) states, irrespective of the command log length.

\subsection{The Shared Log on Blob Storage}

\system aims to operate with minimal setup, preferably on top of off-the-shelf cloud storage, without employing VM-based shared log systems~\cite{lazylog,corfudb,balakrishnan2013tango}.
\system achieves this goal by implementing the shared log atop of existing blob storage.

\mypara{Storage model}
Storage should offer the following operations: (1) create a blob (failing if it already exists), (2) \emph{atomically append} to a blob, (3) read from a blob, and (3) get the size of a blob. 
A blob is finite in size; the size limit is predefined.
\system models each blob as a \emph{segment}. 
\system can support a new storage system by extending the segment with a storage-specific implementation, e.g., a Redis-backed segment in place of an Azure blob.

\mypara{Infinite log on finite blobs}
In \ssmr, both the command log and the state log are instances of the shared log abstraction. 
\system implements the shared log through segments.
That is, the shared log consists of one or more ordered segments, and each segment contains one or more commands.
A new segment is added if the last one fills up. 
New commands append to the last segment, called the \emph{tail segment}.

\looseness=-1 The overall structure (i.e., a list of segments and their order) and their physical mapping (to blobs) are recorded in the special \emph{meta segment}. The meta segments record the history of new segment additions.

\looseness=-1 Any \system library (running within an application client) can add a segment as the new tail.
To do so, it first creates a new segment with a unique name (e.g., using a GUID), and then updates the meta segment.
\system loads the meta segment at startup and checks for updates, particularly when existing segments fill up.
For the meta segment itself, we use a static naming scheme of fixed-size segments (e.g., \emph{meta0}, \emph{meta1}, and so on).

\subsection{Optimizations}

To mitigate the latency and cost of interactions with external storage, \system{} introduces two key optimizations.

\mypara{Speculative Reads} When a client reads a key, the latest value may reside either in the command log or in the state log.
Naively, \system{} can first read the command log; and then read a snapshot from the state log if it does not find a value for the key.
This requires at least two RTTs to cloud storage to serve a linearizable read request.
Instead, we can \textit{speculatively} read from the snapshot, assuming the key is not in the command log.
We discard the value from the snapshot if a new command is found for the key.
Reading the command log and materialized state concurrently reduces the latency observed by each request.

\mypara{Batched Reads and Writes} For higher throughput and lower storage overhead, \system{} uses an optimization called a \textit{bus-stand}~\cite{balakrishnan2024taming}: each client has a single thread that periodically synchronizes with the log.
This synchronization checks the tail of the command log, plays commands back, updates local state, and serves pending read requests.
The thread ensures that there is only one outstanding tail check to the command log.
As a result, incoming reads have to wait until the next ``bus'' leaves the bus-stand for cloud storage; in the worst case, they can just miss it and have to wait for one extra RTT.
The tail offset read by the synchronization becomes the linearization point for the reads assigned to it (or equivalently, the snapshot for SI or SS transactions).
To rate-limit checks, the thread enforces some minimum time period between subsequent syncs (e.g., 10ms).
Similarly, \system{} also uses a conventional batching thread for writes.

\begin{figure*}[t]
\pgfplotsset{
    styleEtcd/.style={
        color=etcd,
        mark=triangle*,
        mark size=1.5pt,
        mark options={fill=etcd, thick},
        very thick,
    },
    styleSlsmr/.style={
        color=slsmr,
        mark=diamond*,
        mark size=1.5pt,
        mark options={fill=slsmr, thick},
        very thick,
    },
    styleAirindex/.style={
        color=AirIndex,
        mark=*, 
        mark size=1pt,
        mark options={fill=AirIndex, thick}, 
        very thick,
    }
}

    \captionsetup[sub]{font=footnotesize,skip=2pt,justification=centering,singlelinecheck=false}
    \centering
    \pgfplotslegendfromname{throughputlegend}\par\smallskip
    \makebox[\linewidth][r]{%
    \begin{minipage}{\dimexpr\linewidth-7mm\relax}
    \subcaptionbox{Insert Only \\ (100\%-inserts)\label{fig:workload_insert}}[0.19\linewidth]{%
        \begin{tikzpicture}[trim axis left,trim axis right]
    \begin{axis}[name=throughputpanel,at={(0,0)},anchor=south west,
        width=24mm,
        height=22mm,
        scale only axis,
        grid=major,
        grid style={dashed, gray!30},
        xlabel={Throughput (ops/s)},
        ylabel={P99 Latency (ms)},
        scaled x ticks=false,
        tick label style={font=\scriptsize},
        label style={font=\small},
        title style={font=\small},
        ymin=0,
        ymax=1700,
        ytick={0, 500, 1000, 1500, 1700},
        xmin=0,
        xmax={15000},
        xtick={0, 5000, 10000, 15000},
        xticklabels={0, 5K, 10K, 15K},
        legend style={
            at={(-0.2, 1.1)}, 
            anchor=south west, 
            draw=none, 
            fill=none, 
            legend columns=3,
            column sep=5pt,
            /tikz/every even column/.append style={column sep=10pt},
            font=\small,
        },
        legend to name=throughputlegend,
        legend cell align=left,
    ]

        \addplot+[
            styleEtcd
        ] table [
            x=thpt, y=lat, col sep=comma
        ] {sections/figures/tikz_export/load/etcd.csv};
        \addlegendentry{etcd (conventional SMR)}

        \addplot+[
            styleSlsmr
        ] table [
            x=thpt, y=lat, col sep=comma
        ] {sections/figures/tikz_export/load/slsmr.csv};
        \addlegendentry{SLSMR on Azure Blob (disaggregated SMR)}

        \addplot+[
            styleAirindex
        ] table [
            x=thpt, y=lat, col sep=comma
        ] {sections/figures/tikz_export/load/azure.csv};
        \addlegendentry{\ssmr on Azure Blob (\textbf{ours}) }

    \end{axis}
\pgfresetboundingbox
    \path[use as bounding box] (0,-9mm) rectangle (24mm,24mm);
\end{tikzpicture}%
    }\hfill
    \subcaptionbox{Update Heavy \\ (50\%-reads / 50\%-updates)\label{fig:workloada}}[0.19\linewidth]{%
        \begin{tikzpicture}[trim axis left,trim axis right]
    \begin{axis}[name=throughputpanel,at={(0,0)},anchor=south west,
        width=24mm,
        height=22mm,
        scale only axis,
        grid=major,
        grid style={dashed, gray!30},
        xlabel={Throughput (ops/s)},
        scaled x ticks=false,
        tick label style={font=\scriptsize},
        label style={font=\small},
        title style={font=\small},
        legend style={
            at={(-0.8, 1.1)}, 
            anchor=south west, 
            draw=none, 
            fill=none, 
            legend columns=3,
            column sep=5pt,
            /tikz/every even column/.append style={column sep=20pt},
            font=\small,
        },
        legend cell align=left,
        ymin=0,
        ymax=1200,
        ytick={0, 500, 1000, 1200},
        xtick={0, 10000, 20000, 30000},
        xticklabels={0, 10K, 20K, 30K},
        xmin=0,
        xmax=30000,
    ]

        \addplot+[
            styleEtcd
        ] table [
            x=thpt, y=lat, col sep=comma
        ] {sections/figures/tikz_export/workloada/etcd.csv};

        \addplot+[
            styleSlsmr
        ] table [
            x=thpt, y=lat, col sep=comma
        ] {sections/figures/tikz_export/workloada/slsmr.csv};

        \addplot+[
            styleAirindex
        ] table [
            x=thpt, y=lat, col sep=comma
        ] {sections/figures/tikz_export/workloada/azure.csv};

    \end{axis}
\pgfresetboundingbox
    \path[use as bounding box] (0,-9mm) rectangle (24mm,24mm);
\end{tikzpicture}%
    }\hfill
    \subcaptionbox{Read Heavy \\ (95\%-reads / 5\%-updates)\label{fig:workloadb}}[0.19\linewidth]{%
        \begin{tikzpicture}[trim axis left,trim axis right]
    \begin{axis}[name=throughputpanel,at={(0,0)},anchor=south west,
        width=24mm,
        height=22mm,
        scale only axis,
        grid=major,
        grid style={dashed, gray!30},
        xlabel={Throughput (ops/s)},
        scaled x ticks=false,
        tick label style={font=\scriptsize},
        label style={font=\small},
        title style={font=\small},
        legend style={
            at={(0.0,1.1)}, 
            anchor=south west, 
            draw=none, 
            fill=none, 
            legend columns=3,
            column sep=1.5em,
            font=\scriptsize,
        },
        ymin=0,
        ymax=800,
        ytick={0, 200, 400, 600, 800},
        xmin=0,
        xmax=100000,
        xtick={0, 20000, 40000, 60000, 80000, 100000},
        xticklabels={0, 20K, 40K, 60K, 80K, 100K},
    ]

        \addplot+[
            styleEtcd
        ] table [
            x=thpt, y=lat, col sep=comma
        ] {sections/figures/tikz_export/workloadb/etcd.csv};

        \addplot+[
            styleSlsmr
        ] table [
            x=thpt, y=lat, col sep=comma
        ] {sections/figures/tikz_export/workloadb/slsmr.csv};

        \addplot+[
            styleAirindex
        ] table [
            x=thpt, y=lat, col sep=comma
        ] {sections/figures/tikz_export/workloadb/azure.csv};

    \end{axis}
\pgfresetboundingbox
    \path[use as bounding box] (0,-9mm) rectangle (24mm,24mm);
\end{tikzpicture}%
    }\hfill
    \subcaptionbox{Read Only \\ (100\%-reads)\label{fig:workloadc}}[0.19\linewidth]{%
        \begin{tikzpicture}[trim axis left,trim axis right]
    \begin{axis}[name=throughputpanel,at={(0,0)},anchor=south west,
        width=24mm,
        height=22mm,
        scale only axis,
        grid=major,
        grid style={dashed, gray!30},
        xlabel={Throughput (ops/s)},
        scaled x ticks=false,
        tick label style={font=\scriptsize},
        label style={font=\small},
        title style={font=\scriptsize},
        ymin=0,
        ymax=400,
        ytick={0, 100, 200, 300, 400},
        xmin=0,
        xmax=200000,
        xtick={0, 50000, 100000, 150000, 200000},
        xticklabels={0, 50K, 100K, 150K, 200K},
    ]

        \addplot+[
            styleEtcd
        ] table [
            x=thpt, y=lat, col sep=comma
        ] {sections/figures/tikz_export/workloadc/etcd.csv};

        \addplot+[
            styleSlsmr
        ] table [
            x=thpt, y=lat, col sep=comma
        ] {sections/figures/tikz_export/workloadc/slsmr.csv};

        \addplot+[
            styleAirindex
        ] table [
            x=thpt, y=lat, col sep=comma
        ] {sections/figures/tikz_export/workloadc/azure.csv};

    \end{axis}
\pgfresetboundingbox
    \path[use as bounding box] (0,-9mm) rectangle (24mm,24mm);
\end{tikzpicture}%
    }\hfill
    \subcaptionbox{Read-Modify-Write \\ (50\%-RMW / 50\%-reads)\label{fig:workloadf}}[0.19\linewidth]{%
        \begin{tikzpicture}[trim axis left,trim axis right]
    \begin{axis}[name=throughputpanel,at={(0,0)},anchor=south west,
        width=24mm,
        height=22mm,
        scale only axis,
        grid=major,
        grid style={dashed, gray!30},
        xlabel={Throughput (ops/s)},
        scaled x ticks=false,
        tick label style={font=\scriptsize},
        label style={font=\small},
        title style={font=\scriptsize},
        ymin=0,
        ymax=2500,
        ytick={0, 1000, 2000, 2500},
        xmin=0,
        xmax=60000,
        xtick={0, 20000, 40000, 60000},
        xticklabels={0, 20K, 40K, 60K},
    ]

        \addplot+[
            styleEtcd
        ] table [
            x=thpt, y=lat, col sep=comma
        ] {sections/figures/tikz_export/workloadf/etcd.csv};

        \addplot+[
            styleSlsmr
        ] table [
            x=thpt, y=lat, col sep=comma
        ] {sections/figures/tikz_export/workloadf/slsmr.csv};

        \addplot+[
            styleAirindex
        ] table [
            x=thpt, y=lat, col sep=comma
        ] {sections/figures/tikz_export/workloadf/azure.csv};

    \end{axis}
\pgfresetboundingbox
    \path[use as bounding box] (0,-9mm) rectangle (24mm,24mm);
\end{tikzpicture}%
    }
    \end{minipage}%
    }

    \caption{Latency-Throughput comparison with YCSB workloads. \system{}, an instance of \ssmr, achieves competitive performance compared to etcd (conventional SMR) and SLSMR (shared log-based disaggregated SMR) across all workload types.}
    \label{fig:throughput-latency}
\end{figure*}
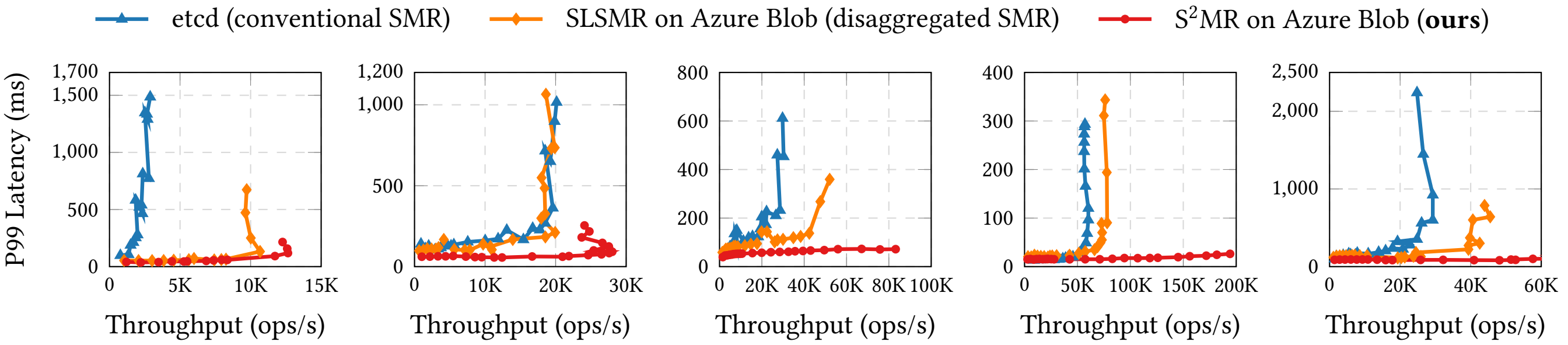

\section{Evaluation} 
\label{sec:eval}

Our evaluation aims to answer the following questions:

\begin{itemize}[leftmargin=12pt, rightmargin=0pt]
    \item \textbf{Performance}: Can a \ssmr-based system offer competitive throughput-latency when compared to the systems following more conventional SMR architecture? (\cref{sec:eval:throughput})
    \item \textbf{Availability}: Does eliminating servers preserve uptime during node, client, and storage failures? (\cref{sec:eval:availability})
    \item \textbf{Impact of Storage}: How does the storage performance affect \ssmr performance? (\cref{sec:eval:storage})
    \item \textbf{Cost-Efficiency}: Can \ssmr lower infrastructure costs, even considering client costs? (\cref{sec:eval:cost})
    \item \textbf{Scalability}: Can \ssmr offer higher overall throughput when more clients interact with the shared storage? (\cref{sec:eval:client})
\end{itemize}

\noindent
In the remainder of this section,
we describe a common evaluation setup
    and then present evaluation results.

\subsection{Setup}
\label{sec:eval:setup}

We evaluate \ssmr{} through \system{}, its transactional key-value store implementation. All systems in the comparison provide strict serializable execution.

\begin{itemize}[leftmargin=*]

\item \textbf{\emph{etcd}}~\cite{etcd} is a Raft-based strongly consistent, distributed key-value store. We enable strict serializable execution.

\item \textbf{\emph{Shared Log SMR (SLSMR)}} is a shared-log-based distributed multi-master database. 
SLSMR uses Azure Blob Storage to implement the shared-log abstraction, i.e., the same approach as \system{}. 
Each of SLSMR's masters (re-)constructs the state by replaying commands (in the shared log) or loading snapshots.
The state is kept in memory. 
SLSMR is configured to use strict serializability.

\item \textbf{\emph{\system{}}} is our \ssmr implementation. It offers the transactional key-value store interface. 
To implement the shared log abstraction, \system uses Azure Blob (for \cref{sec:eval:throughput}, \cref{sec:eval:availability}, and \cref{sec:eval:cost}) and additionally uses Redis (\cref{sec:eval:storage}).
It is set to use strict serializability for all our evaluations.

\end{itemize}

\smallpara{Workloads}
We use workloads from the YCSB benchmark \cite{ycsb}. We test four workload types: insert-only; update-heavy (50/50 update/read); read-heavy (5/95 update/read); read-only; and read–modify–write (50/50 RMW/read). All experiments operate on 1\,GB data containing one million key–value pairs. Each pair is 1\,KB in size. Unless otherwise noted, we set $\theta = 0$; keys are accessed with equal probability.
We choose 1\,GB to reflect the scale of coordination services such as etcd, which enforces a default storage limit of 2\,GB.

\smallpara{Environments}
We use Azure~\cite{azcloud} within a single virtual network. The baseline systems (i.e., etcd and SLSMR) run on three-node clusters of Standard D4\_v3 VMs (4 vCPUs, 8 GiB memory). Clients run on individual instances of the same type. All VMs run Ubuntu 20.04 on standard HDDs. Each VM uses a single standard Azure virtual NIC with IPv4 networking and accelerated networking disabled, capping aggregate bandwidth at roughly 2\,Gbps per VM.

\smallpara{Storage}
We evaluate several storage backends for \system.
First, we use Azure Blob~\cite{azure-blob} for the primary performance and cost analysis.
Second, to demonstrate \system’s extensibility, we also employ a replicated Redis cluster (with our own sequencer) to represent a low-latency shared-log storage.

\subsection{\ssmr Delivers High Throughput}
\label{sec:eval:throughput}

We examine \system{}'s throughput–latency against existing SMR systems, studying whether \system{}, running on external storage, can offer low-latency operations as load increases.

\smallpara{Approach}
We compare \system{} against etcd and SLSMR using throughput–latency curves collected after a one-minute warm-up. The five YCSB workloads comprise four single-operation types and read-modify-write (RMW), whose multi-operation commands commit only when conflict-free under strict serializability. We use the 1\,GB dataset (cold start is studied separately in \cref{sec:eval:availabilty:coldstart}); insert-only (\cref{fig:workload_insert}) starts from an empty database. Each workload runs three times; we report average throughput and P99 latency over a one-minute window, issued from three client VMs and controlled by the number of in-flight operations.

\smallpara{Results}
\cref{fig:throughput-latency} shows P99 latency as a function of throughput. Overall, \system{} reaches higher peak throughput than the baselines.
For read-heavy and read-only workloads (\cref{fig:workloadb,fig:workloadc}), \system{} achieves up to 
2.5$\times$ the throughput of SLSMR and 4$\times$ that of etcd. At lower offered load (<50K ops/s), etcd reports lower latency.
For workloads involving writes (\cref{fig:workload_insert}), both \system{} and SLSMR attain higher throughput than etcd. SLSMR exhibits somewhat higher latency than \system{} at comparable throughput levels.
For the RMW workload (\cref{fig:workloadf}), the uniform key distribution yields low contention, \system{} achieves up to 3$\times$ the throughput of etcd. 

\begin{figure}[t]
    \centering
    \captionsetup{skip=5pt}
    \pgfmathsetmacro{\plotwidth}{55}
    \pgfmathsetmacro{\plotheight}{14}
    \pgfplotslegendfromname{availabilitylegend}\par\smallskip
    \makebox[\linewidth][c]{\begin{tikzpicture}[trim axis left, trim axis right]
    \pgfplotstableread[col sep=comma]{sections/figures/tikz_export/etcd_availability/etcd_availability_timeseries.csv}\datatable

    \pgfplotsset{
        timelineStyle/.style={
            width=\plotwidth mm,
            height=\plotheight mm,
            scale only axis,
            at={(0,0)},anchor=south west,
            xmin=0,
            xmax=180,
            clip=false 
        }
    }

    \begin{axis}[
        timelineStyle,
        grid=major,
        grid style={dashed, gray!30},
        axis y line*=left,
        ylabel={P99 Latency (ms)},
        xlabel style={font=\small},
        ylabel style={StrongBlue, font=\scriptsize},
        yticklabel style={StrongBlue, font=\scriptsize, text width=2.5em, align=right},
        xticklabel style={font=\scriptsize},
        xticklabels=\empty,
        ymin=0,
        clip=false,
    ]
        \addplot[color=StrongBlue, thick] table [x index=0, y index=2] {\datatable};
        \label{plot:latency}
        
        \draw [black, dashed, thick, opacity=0.5] (axis cs:29.98, \pgfkeysvalueof{/pgfplots/ymin}) -- (axis cs:29.98, \pgfkeysvalueof{/pgfplots/ymax} + 0.1);
        \draw [black, dashed, thick, opacity=0.5] (axis cs:61.01, \pgfkeysvalueof{/pgfplots/ymin}) -- (axis cs:61.01, \pgfkeysvalueof{/pgfplots/ymax} + 0.1);
        \draw [black, dashed, thick, opacity=0.5] (axis cs:92.46, \pgfkeysvalueof{/pgfplots/ymin}) -- (axis cs:92.46, \pgfkeysvalueof{/pgfplots/ymax} + 0.1);
    \end{axis}

    \begin{axis}[
        timelineStyle,
        axis y line*=right,
        axis x line=none,
        ylabel={Thpt (ops/s)},
        ylabel style={BrightRed, font=\scriptsize},
        yticklabel style={BrightRed, font=\scriptsize},
        ymin=0,
        ymax=1000,
        ytick={0, 500, 1000},
        legend to name=availabilitylegend,
        legend columns=2,
        legend image post style={opacity=1},
        legend cell align=left,
        legend style={draw=none, fill=none, font=\footnotesize, column sep=1em},
    ]
        \addlegendimage{/pgfplots/refstyle=plot:latency}\addlegendentry{P99 Latency}
        
        \addplot[color=BrightRed, thick, opacity=0.5] table [x index=0, y index=1] {\datatable};
        \addlegendentry{Throughput}
        
        \node[anchor=south west, font=\footnotesize\bfseries, inner sep=1.5pt, xshift=-5mm, yshift=3pt]
            at (rel axis cs:0,1) {etcd};

        \node[inner sep=1.5pt, anchor=south, font=\scriptsize, yshift=3pt] 
            at (axis cs:29.98, \pgfkeysvalueof{/pgfplots/ymax}) {1st};
            
        \node[inner sep=1.5pt, anchor=south, font=\scriptsize, yshift=3pt] 
            at (axis cs:61.01, \pgfkeysvalueof{/pgfplots/ymax}) {2nd};
            
        \node[inner sep=1.5pt, anchor=south, font=\scriptsize, yshift=3pt]
            at (axis cs:92.46, \pgfkeysvalueof{/pgfplots/ymax}) {3rd};
    \end{axis}
\pgfresetboundingbox
    \path[use as bounding box] (0,-1mm) rectangle (55mm,18mm);
\end{tikzpicture}\unskip}\par
    \makebox[\linewidth][c]{\begin{tikzpicture}[trim axis left, trim axis right]
    \pgfplotstableread[col sep=comma]{sections/figures/tikz_export/slsmr_availability/slsmr_availability_timeseries.csv}\datatable

    \pgfplotsset{
        timelineStyle/.style={
            width=\plotwidth mm,
            height=\plotheight mm,
            scale only axis,
            at={(0,0)},anchor=south west,
            xmin=0,
            xmax=180,
            clip=false
        }
    }

    \begin{axis}[
        timelineStyle,
        xlabel={Time (s)},
        grid=major,
        grid style={dashed, gray!30},
        axis y line*=left,
        ylabel={P99 Latency (ms)},
        xlabel style={font=\small},
        ylabel style={StrongBlue, font=\scriptsize},
        yticklabel style={StrongBlue, font=\scriptsize, text width=2.5em, align=right},
        xticklabel style={font=\scriptsize},
        ymin=0,
    ]
        \addplot[color=StrongBlue, thick] table [x index=0, y index=2] {\datatable};
        \label{plot:latency_slsmr}

        \draw [black, dashed, thick, opacity=0.5] (axis cs:29.37, \pgfkeysvalueof{/pgfplots/ymin}) -- (axis cs:29.37, \pgfkeysvalueof{/pgfplots/ymax} + 0.1);
        \draw [black, dashed, thick, opacity=0.5] (axis cs:60.86, \pgfkeysvalueof{/pgfplots/ymin}) -- (axis cs:60.86, \pgfkeysvalueof{/pgfplots/ymax} + 0.1);
        \draw [black, dashed, thick, opacity=0.5] (axis cs:92.43, \pgfkeysvalueof{/pgfplots/ymin}) -- (axis cs:92.43, \pgfkeysvalueof{/pgfplots/ymax} + 0.1);
        \draw [black, dashed, thick, opacity=0.5] (axis cs:123.52, \pgfkeysvalueof{/pgfplots/ymin}) -- (axis cs:123.52, \pgfkeysvalueof{/pgfplots/ymax} + 0.1);
        \draw [black, dashed, thick, opacity=0.5] (axis cs:155.3, \pgfkeysvalueof{/pgfplots/ymin}) -- (axis cs:155.3, \pgfkeysvalueof{/pgfplots/ymax} + 0.1);
    \end{axis}

    \begin{axis}[
        timelineStyle,
        axis y line*=right,
        axis x line=none,
        ylabel={Thpt (ops/s)},
        ylabel style={BrightRed, font=\scriptsize},
        yticklabel style={BrightRed, font=\scriptsize},
        ymin=0,
        ymax=1000,
        ytick={0, 500, 1000},
    ]

        \addplot[color=BrightRed, thick] table [x index=0, y index=1] {\datatable};

        \node[anchor=south west, font=\footnotesize\bfseries, inner sep=1.5pt, xshift=-5mm, yshift=3pt]
            at (rel axis cs:0,1) {SLSMR};

        \node[inner sep=1.5pt, anchor=south, font=\scriptsize, yshift=3pt]
            at (axis cs:29.37, \pgfkeysvalueof{/pgfplots/ymax}) {1st};

        \node[inner sep=1.5pt, anchor=south, font=\scriptsize, yshift=3pt]
            at (axis cs:60.86, \pgfkeysvalueof{/pgfplots/ymax}) {2nd};

        \node[inner sep=1.5pt, anchor=south, font=\scriptsize, yshift=3pt]
            at (axis cs:92.43, \pgfkeysvalueof{/pgfplots/ymax}) {3rd};

        \node[inner sep=1.5pt, anchor=south, font=\scriptsize, yshift=3pt]
            at (axis cs:123.52, \pgfkeysvalueof{/pgfplots/ymax}) {4th};

        \node[inner sep=1.5pt, anchor=south, font=\scriptsize, yshift=3pt]
            at (axis cs:155.3, \pgfkeysvalueof{/pgfplots/ymax}) {5th};
    \end{axis}
\pgfresetboundingbox
    \path[use as bounding box] (0,-9mm) rectangle (55mm,18mm);
\end{tikzpicture}\unskip}\par
    \caption{Server failures' impact on availability. Each vertical dashed line indicates a node failure.
    \textbf{(top)} etcd becomes unavailable after the 3rd failure (loses quorum).
    \textbf{(bottom)} SLSMR tolerates 4 failures but crashes on the 5th (no remaining replicas).
    \system{} is unaffected by server failures by design, thus is omitted in this figure.}
    \label{fig:availability}
\end{figure}
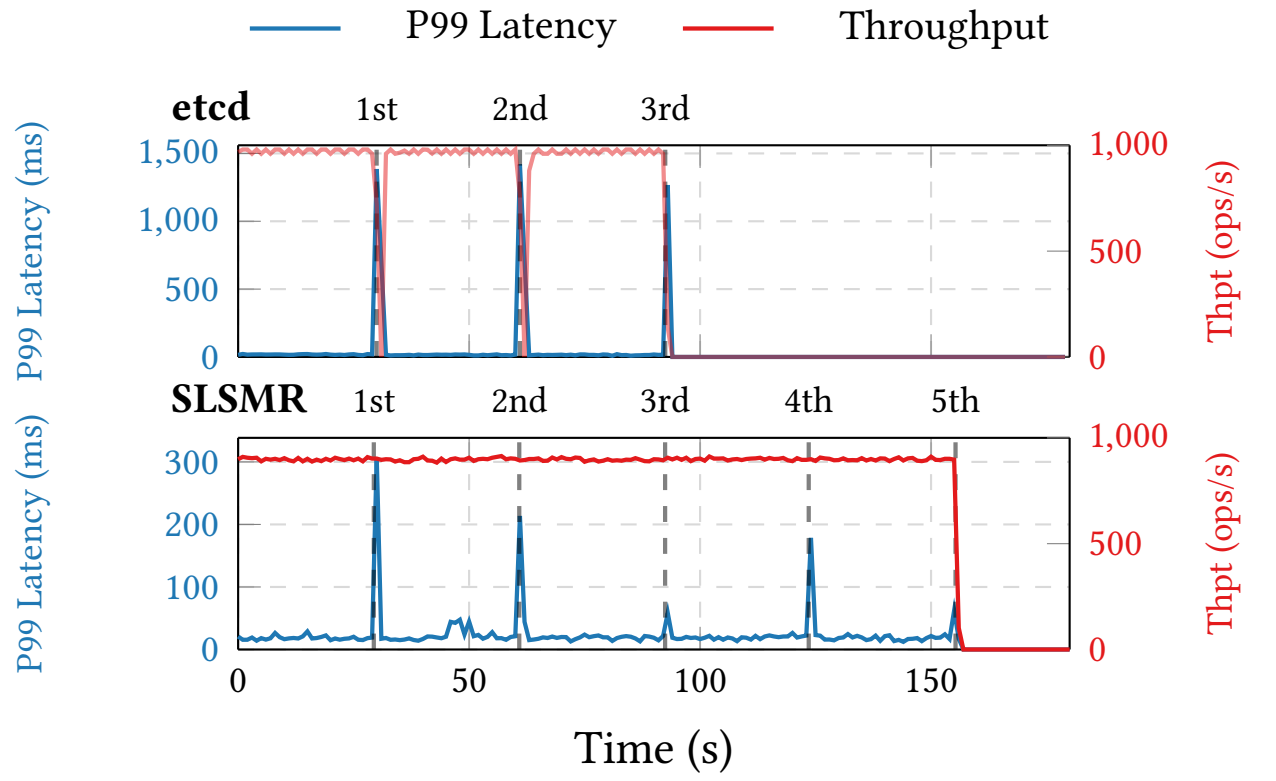

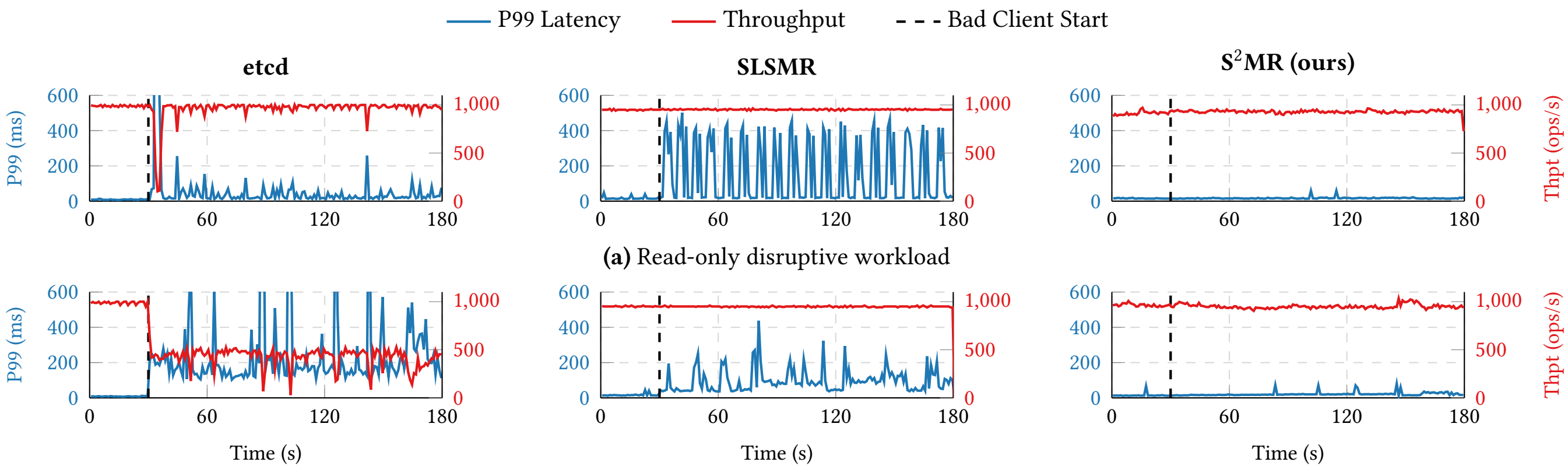
\begin{figure*}[t]
\centering
\captionsetup{skip=5pt}
\captionsetup[sub]{font=footnotesize,skip=2pt,belowskip=0pt,justification=centering}
\pgfplotsset{
    badclientaxis/.style={
        width=40mm,height=12mm,scale only axis,
        anchor=south west,xmin=0,xmax=180,ymin=0,
        xtick={0,60,120,180},scaled ticks=false,
        tick label style={font=\scriptsize},
        label style={font=\scriptsize},
        title style={font=\footnotesize\bfseries,yshift=-3pt},
    },
    badclientlatency/.style={badclientaxis,
        axis y line*=left,axis x line*=bottom,
        grid=major,grid style={dashed,gray!30},
        ymax=600,ytick={0,200,400,600},
        ylabel style={BlueColor,font=\scriptsize},
        yticklabel style={BlueColor,font=\scriptsize},
    },
    badclientthroughput/.style={badclientaxis,
        axis y line*=right,axis x line=none,
        ymax=1100,ytick={0,500,1000},
        ylabel style={RedColor,font=\scriptsize},
        yticklabel style={RedColor,font=\scriptsize},
    }
}
{\footnotesize
\tikz[baseline=-0.5ex]\draw[BlueColor,thick](0,0)--(5mm,0);\ P99 Latency\qquad
\tikz[baseline=-0.5ex]\draw[RedColor,thick](0,0)--(5mm,0);\ Throughput\qquad
\tikz[baseline=-0.5ex]\draw[black,dashed,thick](0,0)--(5mm,0);\ Bad Client Start
}\par\smallskip
\noindent\begin{subfigure}[b]{\linewidth}
\centering
\begin{tikzpicture}
\begin{axis}[badclientlatency,at={(0mm,0)},title={etcd},xlabel={},ylabel={P99 (ms)}]
    \addplot[BlueColor,thick] table[x=time,y=p99_latency,col sep=comma]{sections/figures/tikz_export/bad_client_read_only/etcd.csv};
    \draw[black,dashed,thick] (axis cs:30,0)--(axis cs:30,600);
\end{axis}
\begin{axis}[badclientthroughput,at={(0mm,0)},ylabel={}]
    \addplot[RedColor,thick] table[x=time,y=throughput,col sep=comma]{sections/figures/tikz_export/bad_client_read_only/etcd.csv};
\end{axis}
\begin{axis}[badclientlatency,at={(58mm,0)},title={SLSMR},xlabel={},ylabel={}]
    \addplot[BlueColor,thick] table[x=time,y=p99_latency,col sep=comma]{sections/figures/tikz_export/bad_client_read_only/slsmr.csv};
    \draw[black,dashed,thick] (axis cs:30,0)--(axis cs:30,600);
\end{axis}
\begin{axis}[badclientthroughput,at={(58mm,0)},ylabel={}]
    \addplot[RedColor,thick] table[x=time,y=throughput,col sep=comma]{sections/figures/tikz_export/bad_client_read_only/slsmr.csv};
\end{axis}
\begin{axis}[badclientlatency,at={(116mm,0)},title={\ssmr (ours)},xlabel={},ylabel={}]
    \addplot[BlueColor,thick] table[x=time,y=p99_latency,col sep=comma]{sections/figures/tikz_export/bad_client_read_only/ssmr.csv};
    \draw[black,dashed,thick] (axis cs:30,0)--(axis cs:30,600);
\end{axis}
\begin{axis}[badclientthroughput,at={(116mm,0)},ylabel={Thpt (ops/s)}]
    \addplot[RedColor,thick] table[x=time,y=throughput,col sep=comma]{sections/figures/tikz_export/bad_client_read_only/ssmr.csv};
\end{axis}
\pgfresetboundingbox
\path[use as bounding box] (-10mm,-4mm) rectangle (166mm,18mm);
\end{tikzpicture}
\caption{Read-only disruptive workload}
\label{fig:bad_client_read_only}
\end{subfigure}\unskip\par\smallskip
\noindent\begin{subfigure}[b]{\linewidth}
\centering
\begin{tikzpicture}
\begin{axis}[badclientlatency,at={(0mm,0)},title={},xlabel={Time (s)},ylabel={P99 (ms)}]
    \addplot[BlueColor,thick] table[x=time,y=p99_latency,col sep=comma]{sections/figures/tikz_export/bad_client_update/etcd.csv};
    \draw[black,dashed,thick] (axis cs:30,0)--(axis cs:30,600);
\end{axis}
\begin{axis}[badclientthroughput,at={(0mm,0)},ylabel={}]
    \addplot[RedColor,thick] table[x=time,y=throughput,col sep=comma]{sections/figures/tikz_export/bad_client_update/etcd.csv};
\end{axis}
\begin{axis}[badclientlatency,at={(58mm,0)},title={},xlabel={Time (s)},ylabel={}]
    \addplot[BlueColor,thick] table[x=time,y=p99_latency,col sep=comma]{sections/figures/tikz_export/bad_client_update/slsmr.csv};
    \draw[black,dashed,thick] (axis cs:30,0)--(axis cs:30,600);
\end{axis}
\begin{axis}[badclientthroughput,at={(58mm,0)},ylabel={}]
    \addplot[RedColor,thick] table[x=time,y=throughput,col sep=comma]{sections/figures/tikz_export/bad_client_update/slsmr.csv};
\end{axis}
\begin{axis}[badclientlatency,at={(116mm,0)},title={},xlabel={Time (s)},ylabel={}]
    \addplot[BlueColor,thick] table[x=time,y=p99_latency,col sep=comma]{sections/figures/tikz_export/bad_client_update/ssmr.csv};
    \draw[black,dashed,thick] (axis cs:30,0)--(axis cs:30,600);
\end{axis}
\begin{axis}[badclientthroughput,at={(116mm,0)},ylabel={Thpt (ops/s)}]
    \addplot[RedColor,thick] table[x=time,y=throughput,col sep=comma]{sections/figures/tikz_export/bad_client_update/ssmr.csv};
\end{axis}
\pgfresetboundingbox
\path[use as bounding box] (-10mm,-9mm) rectangle (166mm,13mm);
\end{tikzpicture}
\caption{Update-heavy disruptive workload}
\label{fig:bad_client_update}
\end{subfigure}\unskip\par
\caption{Impact of disruptive (``bad'') clients on well-behaved (``good'') clients.
Each row shows a different disruptive workload: (a) read-heavy and (b) update-heavy.
Within each row, columns correspond to etcd, SLSMR, and \system{}, respectively.
Each subplot reports two metrics for the Good Client over time: P99 latency (blue, left axis) and throughput (red, right axis).
The vertical dashed line marks the moment when Bad Clients begin issuing high-rate requests.}
\label{fig:bad_client}
\end{figure*}

\smallpara{Discussion}
These trends reflect architectural differences. etcd serves reads locally at low load but its latency rises sharply once the leader becomes CPU-bound; \system{} and SLSMR instead pay a fixed round trip to validate the log tail, which adds latency at low load but does not grow with throughput. This fixed cost stays roughly constant as load rises, whereas etcd's latency degrades exponentially once the leader's CPU saturates with Raft coordination.
\system{}'s latency floor is thus set by the storage round-trip time, trading low-load latency for higher peak throughput and availability; client-side caching can further cut read latency.
For write-heavy workloads, the disaggregated designs avoid Raft's quorum replication and local persistence, sustaining higher throughput: writing directly to the disaggregated shared log removes the leader--follower replication bottleneck, letting write throughput scale with the storage layer rather than a consensus engine. SLSMR's extra network hop raises its tail latency relative to \system{}, and the RMW workload's uniform access yields little contention.

\subsection{\ssmr is More Available}
\label{sec:eval:availability}

We assess \system{}'s availability under three failure classes: application-server failures, cold starts after idle periods, and disruptive clients degrading availability for benign ones.

\subsubsection{Application Server Failure}

We evaluate system availability under application-server failures in comparison with etcd and SLSMR.

\smallpara{Approach}
We deploy five-node clusters for etcd and SLSMR and sequentially fail one node every 30 seconds. A client issues a read-only workload at a target throughput of 1,000 operations per second. Throughput and P99 latency are recorded during the sequence of failures.

\smallpara{Results}
\cref{fig:availability} shows that etcd remains available until three nodes fail, at which point quorum is lost. SLSMR continues to operate with up to four failed nodes, as each remaining node can independently access the shared log. Both systems exhibit latency increases during failure.

\smallpara{Discussion}
etcd and SLSMR rely on servers to process operations, so those servers determine service availability. \system{} instead executes directly on storage; failures of application servers cannot make it unavailable.

\subsubsection{Better Isolation}

We evaluate whether \system{} isolates well-behaved clients from disruptive clients using the same shared-storage deployment.

\smallpara{Approach}
We examine two scenarios: Bad Clients disrupt the system with (1) excessive reads and (2) excessive writes.
In both cases, the Good Client issues a YCSB read-only workload on 1\,GB data with a Zipfian distribution ($\theta=0.99$) at 1,000 operations per second. 
At $t=30$s, we introduce Bad Clients that issue high-throughput workloads with a Uniform distribution. etcd and SLSMR run on three-node clusters using D4\_v3 VMs. For \cref{fig:bad_client_read_only}, we launch three Bad Client VMs targeting 100K ops/s each. For \cref{fig:bad_client_update}, we launch one Bad Client targeting 50k ops/s.

\smallpara{Result}
\cref{fig:bad_client_read_only} shows the Good Client's latency and throughput under read-based disruption. In etcd and SLSMR, latency increased immediately at $t=30$s. etcd experienced a brief throughput drop. \system's latency and throughput stayed nearly identical.
\cref{fig:bad_client_update} shows the Good Client's latency and throughput under update-based disruption. etcd showed large latency swings and significant throughput impact since $t=30$. 
SLSMR  also showed latency degradation, though with lower variance. Its throughput remained stable. \system's latency and throughput showed minimal impact.

\begin{figure*}[t]
    \centering
    \vspace{2mm}
    \begin{subfigure}[b]{0.35\textwidth}
        \centering
        \def\vm{45.84}

\pgfplotstableread[row sep=\\, col sep=&]{
    size  & etcdTot & etcdBase & slsmrTot & ssmrTot \\
    100MB & 6.11    & 1.61     & 1.80     & 0.85    \\
    500MB & 21.97   & 3.18     & 8.12     & 1.68    \\
    1GB   & 35.70   & 5.81     & 18.09    & 2.37    \\
    5GB   & 232.55  & 25.79    & 21.16    & 6.07    \\
}\coldstartdata

\pgfplotstablegetelem{0}{etcdTot}\of\coldstartdata  \pgfmathsetmacro{\etcdTotA}{\pgfplotsretval}
\pgfplotstablegetelem{0}{slsmrTot}\of\coldstartdata \pgfmathsetmacro{\slsmrTotA}{\pgfplotsretval}
\pgfplotstablegetelem{0}{ssmrTot}\of\coldstartdata  \pgfmathsetmacro{\ssmrTotA}{\pgfplotsretval}
\pgfplotstablegetelem{0}{etcdBase}\of\coldstartdata \pgfmathsetmacro{\etcdBaseA}{\pgfplotsretval}

\pgfplotstablegetelem{1}{etcdTot}\of\coldstartdata  \pgfmathsetmacro{\etcdTotB}{\pgfplotsretval}
\pgfplotstablegetelem{1}{slsmrTot}\of\coldstartdata \pgfmathsetmacro{\slsmrTotB}{\pgfplotsretval}
\pgfplotstablegetelem{1}{ssmrTot}\of\coldstartdata  \pgfmathsetmacro{\ssmrTotB}{\pgfplotsretval}
\pgfplotstablegetelem{1}{etcdBase}\of\coldstartdata \pgfmathsetmacro{\etcdBaseB}{\pgfplotsretval}

\pgfplotstablegetelem{2}{etcdTot}\of\coldstartdata  \pgfmathsetmacro{\etcdTotC}{\pgfplotsretval}
\pgfplotstablegetelem{2}{slsmrTot}\of\coldstartdata \pgfmathsetmacro{\slsmrTotC}{\pgfplotsretval}
\pgfplotstablegetelem{2}{ssmrTot}\of\coldstartdata  \pgfmathsetmacro{\ssmrTotC}{\pgfplotsretval}
\pgfplotstablegetelem{2}{etcdBase}\of\coldstartdata \pgfmathsetmacro{\etcdBaseC}{\pgfplotsretval}

\pgfplotstablegetelem{3}{etcdTot}\of\coldstartdata  \pgfmathsetmacro{\etcdTotD}{\pgfplotsretval}
\pgfplotstablegetelem{3}{slsmrTot}\of\coldstartdata \pgfmathsetmacro{\slsmrTotD}{\pgfplotsretval}
\pgfplotstablegetelem{3}{ssmrTot}\of\coldstartdata  \pgfmathsetmacro{\ssmrTotD}{\pgfplotsretval}
\pgfplotstablegetelem{3}{etcdBase}\of\coldstartdata \pgfmathsetmacro{\etcdBaseD}{\pgfplotsretval}

\begin{tikzpicture}
    \begin{axis}[
        width=60mm,
        height=35mm,
        ybar,
        bar width=1.5mm,
        ymin=0, ymax=350,
        ylabel={Latency (s)},
        xtick={1,2,3},
        xticklabels={{etcd}, {SLSMR}, {SSMR}},
        xticklabel style={align=center, yshift=-5mm, font=\scriptsize}, 
        yticklabel style={align=center, font=\scriptsize}, 
        grid=major,
        grid style={dashed, gray!30},
        legend columns=1, 
        legend style={
            at={(1.05, 1.1)}, 
            anchor=north east, 
            font=\scriptsize, 
            draw=black,
            fill=white, fill opacity=1,
        },
        legend cell align=left, 
        enlarge x limits=0.35, 
        clip=false,
        label style={font=\scriptsize},
        tick label style={font=\small},
        area legend,
    ]

    \addplot[fill=colProc, draw=black, bar shift=-0.075\linewidth] coordinates {
        (1, \vm+\etcdTotA) (2, \vm+\slsmrTotA) (3, \ssmrTotA) 
    };
    \addplot[fill=colRest, draw=black, bar shift=-0.075\linewidth] coordinates {
        (1, \vm+\etcdTotA-\etcdBaseA) (2, \vm) (3, 0)
    };
    \addplot[fill=colVM, draw=black, bar shift=-0.075\linewidth] coordinates {
        (1, \vm) (2, \vm) (3, 0)
    };
    \addplot[draw=none, bar shift=-0.075\linewidth, nodes near coords, point meta=explicit symbolic, nodes near coords style={font=\tiny, anchor=south, yshift=1pt}] coordinates {
        (1, \vm+\etcdTotA) [52s] (2, \vm+\slsmrTotA) [47.6s] (3, \ssmrTotA) [0.9s]
    };
    \addplot[draw=none, bar shift=-0.075\linewidth, nodes near coords={100MB}, nodes near coords style={font=\scriptsize, rotate=60, anchor=east, yshift=2pt}] coordinates {(1,0) (2,0) (3,0)};
    \addplot[draw=none, bar shift=-0.075\linewidth, 
    nodes near coords style={font=\footnotesize, anchor=north, yshift=1pt, text=black}] coordinates {(1, \vm)};

    \addplot[fill=colProc, draw=black, bar shift=-0.025\linewidth] coordinates {
        (1, \vm+\etcdTotB) (2, \vm+\slsmrTotB) (3, \ssmrTotB)
    };
    \addplot[fill=colRest, draw=black, bar shift=-0.025\linewidth] coordinates {
        (1, \vm+\etcdTotB-\etcdBaseB) (2, \vm) (3, 0)
    };
    \addplot[fill=colVM, draw=black, bar shift=-0.025\linewidth] coordinates {
        (1, \vm) (2, \vm) (3, 0)
    };
    \addplot[draw=none, bar shift=-0.025\linewidth, nodes near coords, point meta=explicit symbolic, nodes near coords style={font=\tiny, anchor=south, yshift=1pt}] coordinates {
        (1, \vm+\etcdTotB) [67.8s] (2, \vm+\slsmrTotB) [54s] (3, \ssmrTotB) [1.7s]
    };
    \addplot[draw=none, bar shift=-0.025\linewidth, nodes near coords={500MB}, nodes near coords style={font=\scriptsize, rotate=60, anchor=east, yshift=2pt}] coordinates {(1,0) (2,0) (3,0)};
    \addplot[draw=none, bar shift=-0.025\linewidth, 
    nodes near coords style={font=\footnotesize, anchor=north, yshift=1pt, text=black}] coordinates {(1, \vm)};

    \addplot[fill=colProc, draw=black, bar shift=0.025\linewidth] coordinates {
        (1, \vm+\etcdTotC) (2, \vm+\slsmrTotC) (3, \ssmrTotC)
    };
    \addplot[fill=colRest, draw=black, bar shift=0.025\linewidth] coordinates {
        (1, \vm+\etcdTotC-\etcdBaseC) (2, \vm) (3, 0)
    };
    \addplot[fill=colVM, draw=black, bar shift=0.025\linewidth] coordinates {
        (1, \vm) (2, \vm) (3, 0)
    };
    \addplot[draw=none, bar shift=0.025\linewidth, nodes near coords, point meta=explicit symbolic, nodes near coords style={font=\tiny, anchor=south, yshift=1pt}] coordinates {
        (1, \vm+\etcdTotC) [81.5s] (2, \vm+\slsmrTotC) [64s] (3, \ssmrTotC) [2.4s]
    };
    \addplot[draw=none, bar shift=0.025\linewidth, nodes near coords={1GB}, nodes near coords style={font=\scriptsize, rotate=60, anchor=east, yshift=2pt}] coordinates {(1,0) (2,0) (3,0)};
    \addplot[draw=none, bar shift=0.025\linewidth, 
    nodes near coords style={font=\footnotesize, anchor=north, yshift=1pt, text=black}] coordinates {(1, \vm)};

    \addplot[fill=colProc, draw=black, bar shift=0.075\linewidth] coordinates {
        (1, \vm+\etcdTotD) (2, \vm+\slsmrTotD) (3, \ssmrTotD)
    };
    \addplot[fill=colRest, draw=black, bar shift=0.075\linewidth] coordinates {
        (1, \vm+\etcdTotD-\etcdBaseD) (2, \vm) (3, 0)
    };
    \addplot[fill=colVM, draw=black, bar shift=0.075\linewidth] coordinates {
        (1, \vm) (2, \vm) (3, 0)
    };
    \addplot[draw=none, bar shift=0.075\linewidth, nodes near coords, point meta=explicit symbolic, nodes near coords style={font=\tiny, anchor=south, yshift=1pt}] coordinates {
        (1, \vm+\etcdTotD) [278s] (2, \vm+\slsmrTotD) [67s] (3, \ssmrTotD) [6.1s]
    };
    \addplot[draw=none, bar shift=0.075\linewidth, nodes near coords={5GB}, nodes near coords style={font=\scriptsize, rotate=60, anchor=east, yshift=2pt}] coordinates {(1,0) (2,0) (3,0)};
    \addplot[draw=none, bar shift=0.075\linewidth, 
    nodes near coords style={font=\footnotesize, anchor=north, yshift=1pt, text=black}] coordinates {(1, \vm)};

    \legend{} 
    \addlegendimage{ybar, fill=colProc, draw=black}\addlegendentry{Server Startup}
    \addlegendimage{ybar, fill=colRest, draw=black}\addlegendentry{Snapshot Restore}
    \addlegendimage{ybar, fill=colVM, draw=black}\addlegendentry{VM Startup}

    \end{axis}
\end{tikzpicture}

        \vspace{-3mm}
        \caption{Cold start time breakdown}
        \label{fig:cold_start_bar}
    \end{subfigure}
    \hfill
    \begin{subfigure}[b]{0.64\textwidth}
        \centering
        \begin{tikzpicture}
    \pgfplotstableread[col sep=comma]{sections/figures/tikz_export/cold_start/cold_1m.csv}\datatable

    \pgfplotsset{
        timelineStyle/.style={
            width=32mm,
            height=16mm,
            scale only axis,
            xmin=0,
            xmax=300,
            clip=false
        }
    }

    \begin{axis}[
        timelineStyle,
        xlabel={Time (s)},
        grid=major,
        grid style={dashed, gray!30},
        axis y line*=left,
        ylabel={P99 Latency (ms)},
        ylabel style={BlueColor, font=\scriptsize},
        yticklabel style={BlueColor, font=\scriptsize},
        tick label style={font=\scriptsize},
        label style={font=\footnotesize},
        ytick={0, 100, 200, 250},
        yticklabels={0, 100, 200, 250},
        ymin=0,
        ymax=250,
    ]
        \addplot[color=BlueColor, thick] table [x=time, y=p99_latency] {\datatable};
        \label{plot:latency_1m}
    \end{axis}

    \begin{axis}[
        timelineStyle,
        axis y line*=right,
        axis x line=none,
        scaled ticks = false,
        scaled x ticks = false,
        ylabel={Throughput (ops/s)},
        ylabel style={RedColor, font=\scriptsize},
        yticklabel style={RedColor, font=\scriptsize},
        tick label style={font=\scriptsize},
        ymin=0,
        ytick={0, 20000, 40000},
        yticklabels={0, 20K, 40K},
        legend style={
            at={(-0.2,1.1)},
            anchor=south west,
            legend columns=2,
            draw=none,
            fill=white,
            fill opacity=0.8,
            font=\scriptsize,
        },
        /tikz/every even column/.append style={column sep=1em}
    ]
        \addlegendimage{/pgfplots/refstyle=plot:latency_1m}\addlegendentry{P99 Latency}

        \addplot[color=RedColor, thick] table [x=time, y=throughput] {\datatable};
        \addlegendentry{Throughput}
    \end{axis}
\end{tikzpicture}
        \;
        \begin{tikzpicture}
    \pgfplotstableread[col sep=comma]{sections/figures/tikz_export/cold_start/cold_5m.csv}\datatable

    \pgfplotsset{
        timelineStyle/.style={
            width=32mm,
            height=16mm,
            scale only axis,
            xmin=0,
            xmax=300,
            clip=false
        }
    }

    \begin{axis}[
        timelineStyle,
        xlabel={Time (s)},
        grid=major,
        grid style={dashed, gray!30},
        axis y line*=left,
        ylabel={P99 Latency (ms)},
        ylabel style={BlueColor, font=\scriptsize},
        yticklabel style={BlueColor, font=\scriptsize},
        tick label style={font=\scriptsize},
        xlabel style={font=\scriptsize},
        ymin=0,
        ymax=300,
    ]
        \addplot[color=BlueColor, thick] table [x=time, y=p99_latency] {\datatable};
        \label{plot:latency_5m}
    \end{axis}

    \begin{axis}[
        timelineStyle,
        axis y line*=right,
        axis x line=none,
        scaled ticks = false,
        scaled x ticks = false,
        ylabel={Throughput (ops/s)},
        ylabel style={RedColor, font=\scriptsize},
        yticklabel style={RedColor, font=\scriptsize},
        tick label style={font=\scriptsize},
        ymin=0,
        ytick={0, 5000, 10000, 15000},
        yticklabels={0, 5K, 10K, 15K},
        legend style={
            at={(0.99,0.02)},
            anchor=south east,
            legend columns=1,
            draw=none,
            fill=white,
            fill opacity=0.8,
            font=\scriptsize,
        },
    ]
        \addlegendimage{/pgfplots/refstyle=plot:latency_5m}

        \addplot[color=RedColor, thick] table [x=time, y=throughput] {\datatable};
    \end{axis}
\end{tikzpicture}

        \vspace{0mm}
        \caption{\system's cold start performance over time (\textbf{left}: 1GB, \textbf{right}: 5GB)}
        \label{fig:cold_start_timeline}
    \end{subfigure}

    \vspace{-1mm}
    \caption{
    Cold-Start Performance.
\textbf{(a)} Cold-start latency breakdown across dataset sizes (100\,MB--5\,GB). Stacked bars indicate VM startup, snapshot restore, and server initialization.
\textbf{(b)} Throughput and P99 latency during \system{}’s cold start under a Zipfian workload ($\theta=0.99$) with 1\,GB in-memory buffer. 
The left panel (1\,GB data) shows warm-up as the working set enters cache; the right panel (5\,GB data) shows increased performance variance as the data size exceeded the buffer capacity.
    }
    \label{fig:cold_start}
\end{figure*}
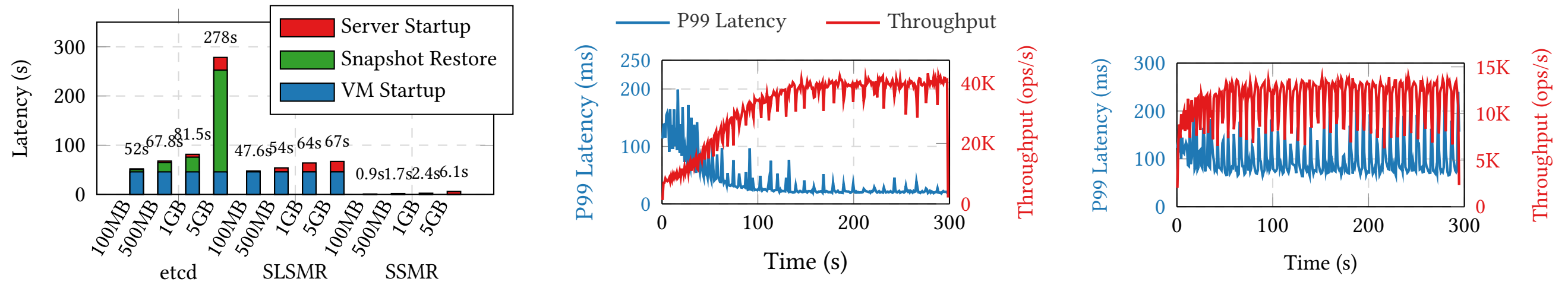

\smallpara{Discussion}
etcd and SLSMR process requests on application servers, so high-rate Bad-Client traffic loads those servers and inflates the Good Client's latency. \system{} executes directly on storage: the Good Client keeps making progress after Bad Clients arrive, with only small latency increases from replaying newly committed updates, while throughput and P99 latency stay stable.

\subsubsection{Faster Cold Start}
\label{sec:eval:availabilty:coldstart}

This section measures cold-start behavior for etcd, SLSMR, and \system{} as the dataset grows.

\smallpara{Approach}
We recorded end-to-end cold-start time, defined as the interval from startup initiation to the first completed transaction. etcd and SLSMR run as three-node clusters and must provision VMs and restore state before serving requests. 
Data sizes were 100\,MB, 500\,MB, 1\,GB, and 5\,GB. 
Each experiment was repeated three times; we report the median.
\system{} initialized directly from storage. It read only the state log indexes and applied the command log tail. 
It deferred full-state reconstruction, filling its cache only as requests accessed data.

\smallpara{Results}
\cref{fig:cold_start_bar} shows etcd and SLSMR slowing sharply as data size grew, since both depend on VM provisioning and full snapshot restoration; \system{} was largely insensitive to data size, touching only metadata and the log tail at startup.
\cref{fig:cold_start_timeline} shows runtime warm-up. With 1\,GB data, throughput rose, and latency stabilized within 150s as the cache filled. 5\,GB data---which exceeds the cache---raised latency variance and lowered throughput as it spilled to storage.

\smallpara{Discussion}
First, cold-start latency grew with dataset size for systems that rebuild full state; \system{} avoided this through storage-centric initialization.
Second, warm-up latency depended on cache size: when the working set fit in memory, the cache stabilized quickly; otherwise, additional storage I/O extended the warm-up period.
Finally, even the system restarting from local disk must load a full snapshot before serving requests, so its recovery time is still impacted by data size (e.g., etcd with 5\,GB in \cref{fig:cold_start_bar}).

\subsubsection{Robustness to Compound Failures}
\label{sec:eval:chaos}

We evaluate the system in an adversarial chaos test.

\smallpara{Approach}
We generated a 100-minute timeline with continuous, artificially injected faults.
For node failures, we configured a 10\% probability of a single VM reboot per minute, and a 1\% probability of a simultaneous, cluster-wide reboot.
Additionally, there was a 10\% probability per minute that a ``bad client'' initiated a sustained, update-heavy workload for one minute.
Finally, we introduced storage instability, where every storage I/O operation had a 0.1\% probability of failure.
These faults stress each architecture differently. VM reboots disrupt only the systems that run operations on dedicated servers (etcd and SLSMR), as \system{} has no servers to reboot. Storage faults disrupt only the systems that execute against the shared log (\system{} and SLSMR), and not etcd, as etcd serves operations from its own nodes' local disks. The bad-client load affects all three systems.

We compare \system{}, etcd, and SLSMR by the fraction of the target client's operations that meet a 100\,ms latency SLA. Throughout the timeline, the target client issues a steady 1K read operations per second. We classify each of its operations as a \emph{Success} (completed within 100\,ms), an \emph{SLA Violation} (completed but slower than 100\,ms), or \emph{Unreachable} (failed to complete because the system was unavailable), and report each category as a percentage of all attempted operations.

\smallpara{Results}
\looseness=-1 Across the 100-minute timeline, we observed 15 individual VM reboots, 2 cluster-wide reboots, and 8 bad-client incidents.
\cref{tab:chaos_results} summarizes the results.
\system{} demonstrates superior resilience, maintaining 99.02\% of operations within the SLA.
With no VM-induced downtime, \system{} records 0.00\% unreachable operations because it decouples compute from state.

\begin{table}[t]

\caption{Aggregate availability and performance under compound failures (VM reboots, bad clients, and storage faults) over a 100-minute timeline.}
    \label{tab:chaos_results}
    
    \centering
    \small
    \begin{tabular}{l c c c}
    \toprule
    \textbf{System} & \textbf{Success} & \textbf{SLA Violation} & \textbf{Unreachable} \\
    & \textbf{(<100ms)} & \textbf{(>100ms)} & \textbf{(Failures)} \\
    \midrule
    etcd      & 89.85\%          & 8.49\%          & 1.66\% \\
   SLSMR     & 92.50\%          & 5.20\%         & 1.28\% \\
    \textbf{\system{}} & \textbf{99.02\%} & \textbf{0.98\%} & \textbf{0.00\%} \\
    \bottomrule
    \end{tabular}
    
\end{table}

\begin{figure*}[t]
\pgfplotsset{
    styleAzureBlob/.style={
        color=AirIndex,
        mark=*,
        mark size=1pt,
        mark options={fill=AirIndex, thick},
        very thick,
    },
    styleRedis/.style={
        color=BlueColor,
        mark=square*,
        mark size=1.5pt,
        mark options={fill=BlueColor, thick},
        very thick,
    }
}

\pgfmathsetmacro{\plotwidth}{24}
\pgfmathsetmacro{\plotheight}{20}

    \captionsetup[sub]{font=footnotesize,belowskip=-2pt,aboveskip=-6pt, justification=centering, singlelinecheck=false}
    \centering
    \begin{subfigure}[b]{0.22\textwidth}
        \begin{tikzpicture}
    \begin{axis}[
        width=\plotwidth mm,
        height=\plotheight mm,
        scale only axis,
        grid=major,
        grid style={dashed, gray!30},
        xlabel={Throughput (ops/s)},
        ylabel={P99 Latency (ms)},
        scaled x ticks=false,
        tick label style={font=\scriptsize},
        label style={font=\small},
        title style={font=\small},
        ymin=0,
        xmin=0,
        xtick={0, 10000, 20000},
        xticklabels={0, 10K, 20K},
        xmax=20000,
        legend style={
            at={(-0.2, 1.05)},
            anchor=south west,
            draw=none,
            fill=none,
            legend columns=2,
            column sep=5pt,
            /tikz/every even column/.append style={column sep=10pt},
            font=\small,
        },
        legend cell align=left,
    ]

        \addplot+[
            styleRedis
        ] table [
            x=thpt, y=lat, col sep=comma
        ] {sections/figures/tikz_export/load/redis.csv};
        \addlegendentry{\ssmr on Redis}

        \addplot+[
            styleAzureBlob
        ] table [
            x=thpt, y=lat, col sep=comma
        ] {sections/figures/tikz_export/load/azure.csv};
        \addlegendentry{\ssmr on Azure Blob}

    \end{axis}
\end{tikzpicture}

        \vspace{-1mm}
        \caption{Insert Only \\ (100\%-inserts)}
        \label{fig:storage_load}
    \end{subfigure}
    \hfill
    \begin{subfigure}[b]{0.19\textwidth}
        \begin{tikzpicture}
    \begin{axis}[
        width=\plotwidth mm,
        height=\plotheight mm,
        scale only axis,
        grid=major,
        grid style={dashed, gray!30},
        xlabel={Throughput (ops/s)},
        scaled x ticks=false,
        tick label style={font=\scriptsize},
        label style={font=\small},
        title style={font=\small},
        ymin=0,
        xmin=0,
        xtick={0, 20000, 40000, 60000},
        xticklabels={0, 20K, 40K, 60k},
        xmax=60000,
    ]

        \addplot+[
            styleAzureBlob
        ] table [
            x=thpt, y=lat, col sep=comma
        ] {sections/figures/tikz_export/workloada/azure.csv};

        \addplot+[
            styleRedis
        ] table [
            x=thpt, y=lat, col sep=comma
        ] {sections/figures/tikz_export/workloada/redis.csv};

    \end{axis}
\end{tikzpicture}

        \vspace{-1mm}
        \caption{Update Heavy \\ (50\%-reads / 50\%-updates)}
        \label{fig:storage_workloada}
    \end{subfigure}
    \hfill
    \begin{subfigure}[b]{0.18\textwidth}
        \begin{tikzpicture}
    \begin{axis}[
        width=\plotwidth mm,
        height=\plotheight mm,
        scale only axis,
        grid=major,
        grid style={dashed, gray!30},
        xlabel={Throughput (ops/s)},
        scaled x ticks=false,
        tick label style={font=\scriptsize},
        label style={font=\small},
        title style={font=\small},
        ymin=0,
        xmin=0,
        xtick={0, 100000, 200000},
        xticklabels={0, 100K, 200K},
        xmax=200000,
    ]

        \addplot+[
            styleAzureBlob
        ] table [
            x=thpt, y=lat, col sep=comma
        ] {sections/figures/tikz_export/workloadb/azure.csv};

        \addplot+[
            styleRedis
        ] table [
            x=thpt, y=lat, col sep=comma
        ] {sections/figures/tikz_export/workloadb/redis.csv};

    \end{axis}
\end{tikzpicture}

        \vspace{-1mm}
        \caption{Read Heavy \\ (95\%-reads / 5\%-updates)}
        \label{fig:storage_workloadb}
    \end{subfigure}
    \hfill
    \begin{subfigure}[b]{0.18\textwidth}
        \begin{tikzpicture}
    \begin{axis}[
        width=\plotwidth mm,
        height=\plotheight mm,
        scale only axis,
        grid=major,
        grid style={dashed, gray!30},
        xlabel={Throughput (ops/s)},
        scaled x ticks=false,
        tick label style={font=\scriptsize},
        label style={font=\small},
        title style={font=\small},
        ymin=0,
        xmin=0,
        xtick={0, 100000, 200000},
        xticklabels={0, 100K, 200K},
        xmax=200000,
    ]

        \addplot+[
            styleAzureBlob
        ] table [
            x=thpt, y=lat, col sep=comma
        ] {sections/figures/tikz_export/workloadc/azure.csv};

        \addplot+[
            styleRedis
        ] table [
            x=thpt, y=lat, col sep=comma
        ] {sections/figures/tikz_export/workloadc/redis.csv};

    \end{axis}
\end{tikzpicture}

        \vspace{-1mm}
        \caption{Read Only \\ (100\%-reads)}
        \label{fig:storage_workloadc}
    \end{subfigure}
    \hfill
    \begin{subfigure}[b]{0.18\textwidth}
        \begin{tikzpicture}
    \begin{axis}[
        width=\plotwidth mm,
        height=\plotheight mm,
        scale only axis,
        grid=major,
        grid style={dashed, gray!30},
        xlabel={Throughput (ops/s)},
        scaled x ticks=false,
        tick label style={font=\scriptsize},
        label style={font=\small},
        title style={font=\small},
        ymin=0,
        xmin=0,
        xtick={0, 50000, 100000},
        xticklabels={0, 50K, 100K},
        xmax=100000,
    ]

        \addplot+[
            styleAzureBlob
        ] table [
            x=thpt, y=lat, col sep=comma
        ] {sections/figures/tikz_export/workloadf/azure.csv};

        \addplot+[
            styleRedis
        ] table [
            x=thpt, y=lat, col sep=comma
        ] {sections/figures/tikz_export/workloadf/redis.csv};

    \end{axis}
\end{tikzpicture}

        \vspace{-1mm}
        \caption{Read-Modify-Write \\ (50\%-RMW / 50\%-reads)}
        \label{fig:storage_workloadf}
    \end{subfigure}

    \caption{
    Impact of storage performance on \ssmr.
Throughput–latency curves with Redis and Azure Blob across five YCSB workloads. 
Each subplot shows P99 latency versus throughput for a distinct workload. }
    \label{fig:storage_comparison}
\end{figure*}
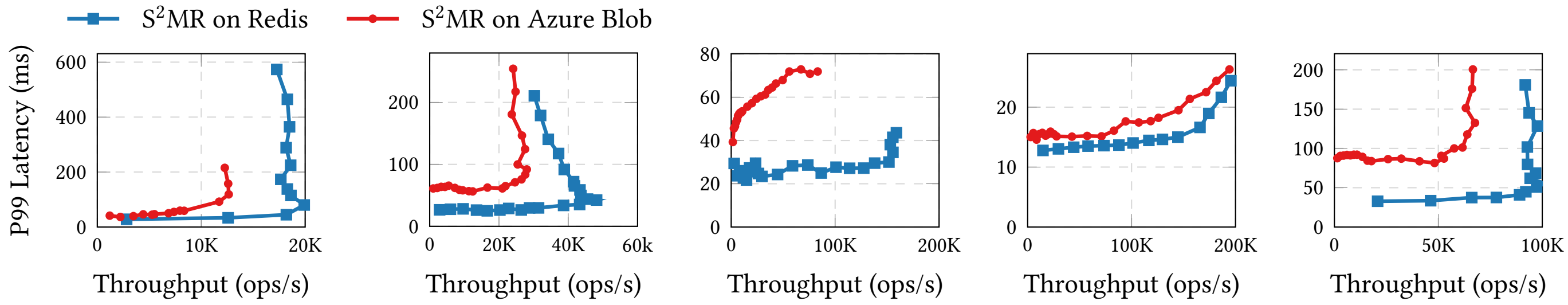 

\subsection{Faster Storage Reduces Latency}
\label{sec:eval:storage}

We study the impact of storage latency on \ssmr{}. 

\smallpara{Approach}
To represent a low-latency shared-log system,
we implemented the shared log interface on top of Redis 
using Lua scripts and locks.
Then, we ran the YCSB benchmarks on 1\,GB data with \system, backed by a three-node Redis deployment (one master, two replicas).

\smallpara{Results}
\cref{fig:storage_comparison} shows that \system{} achieves lower request latency across all workloads when backed by the low-latency Redis storage. The improvement is largest in workloads where \system{} must roll forward new transactions, since these operations depend directly on storage responsiveness: a faster backend shortens each tail check and log-tail read (\cref{fig:storage_load,fig:storage_workloada,fig:storage_workloadb,fig:storage_workloadf}). Across these workloads, Redis lowers \system{}'s latency by 2--45 ms (\cref{fig:storage_workloadc,fig:storage_workloadf}).

\smallpara{Discussion}
\system{} benefits most from faster storage on workloads that frequently process new log entries because each request spends less time fetching and applying committed transactions.
These results also show that \ssmr{} accepts different shared-log backends. Cloud storage and Redis both implement its core abstraction through two operations: \texttt{append} and \texttt{read}.

\subsection{\ssmr is Cost-Efficient}
\label{sec:eval:cost}

We evaluate the operational cost of \system{} against etcd and SLSMR using Azure pricing. 
We used 1\,GB data size and compared two monthly workloads: a \textit{low-load} scenario ($N_{\mathrm{w}}{=}N_{\mathrm{r}}{=}1\mathrm{M}$) and a \textit{high-load} scenario ($N_{\mathrm{w}}{=}N_{\mathrm{r}}{=}100\mathrm{M}$).

\smallpara{Pricing Model}  We use the following Azure costs:
\begin{itemize}
    \item $C_{\mathrm{vm}}$: Monthly cost of a Standard D4\,v3 VM (\$308.06).
    \item $C_{\mathrm{stor}}$: Monthly storage cost per GB (\$0.15).
    \item $C_{\mathrm{mem}}$: Monthly cost of each client's 2GB RAM (\$38.51).
    \item $C_{\mathrm{write}}$: Cost per 10,000 blob write operations (\$0.0228).
    \item $C_{\mathrm{read}}$: Cost per 10,000 blob read operations (\$0.0019).
\end{itemize}

\smallpara{Approach}
\system{} uses client-side caching to reduce latency and cost. This design requires additional client memory (up to 2 GB RAM for a 1 GB database). To evaluate an upper bound on cost, we also analyze a scenario with a 0\% cache hit rate and include the full client memory cost.

\smallpara{Result}
\cref{tab:cost-compact} reports monthly costs. \system{} replaces the fixed expense of dedicated compute nodes with a small client-memory cost ($C_{\mathrm{mem}}$): etcd incurs a constant \$924.18/month and SLSMR needs at least one sequencer VM at \$308.06, both of which \system{} avoids. For 1M ops/mo, \system{} costs \$41.32---about $22\times$ cheaper than etcd and $7.5\times$ than SLSMR---and even in the worst-case 100M-ops workload with no caching it keeps a $3\times$ advantage over etcd.

\begin{table}[t]
\caption{
Monthly cost comparison under Azure pricing. We evaluate a low-load ($1\mathrm{M}$ reads/writes) and a high-load ($100\mathrm{M}$ reads/writes) scenarios. \system{} includes the cost of client memory ($C_{\mathrm{mem}}$) under worst-case 0\% cache hit rate.
}
\label{tab:cost-compact}

\centering
\renewcommand{\arraystretch}{1.2}
\setlength{\tabcolsep}{3pt}
\resizebox{\linewidth}{!}{%
\begin{tabular}{lllll}
\toprule
\textbf{System} & \textbf{Monthly Cost Model} & \textbf{Client} & Low Load & High Load \\
 & & \textbf{Overhead} & 1M R/W & 100M R/W \\
\midrule
etcd
  & $3 C_{\mathrm{vm}}$
  & --
  & \$924.18
  & \$924.18 \\

SLSMR
  & $C_{\mathrm{vm}} {+} C_{\mathrm{stor}}
     {+} N_{\mathrm{w}} (C_{\mathrm{write}} {+} C_{\mathrm{read}})
     {+} N_{\mathrm{r}} C_{\mathrm{read}}$
  & --
  & \$310.87
  & \$574.21 \\

\system{}
  & $C_{\mathrm{mem}} {+} C_{\mathrm{stor}}
     {+} N_{\mathrm{w}} (C_{\mathrm{write}} {+} C_{\mathrm{read}})
     {+} N_{\mathrm{r}} (C_{\mathrm{read}})$
  & \textbf{\$38.51}
  & \textbf{\$41.32}
  & \textbf{\$304.66} \\
\bottomrule
\end{tabular}
}

\end{table}

\subsection{\ssmr Scales Well with More Clients}
\label{sec:eval:client}

We measure how \system{}’s throughput changes as the number of clients increases.
We use YCSB workloads: 1\,GB data; data was fully cached. The number of client VMs varied from 10 to 80. We report aggregate throughput and P99 latency.

\pgfmathsetmacro{\plotwidth}{22}
\pgfmathsetmacro{\plotheight}{15}

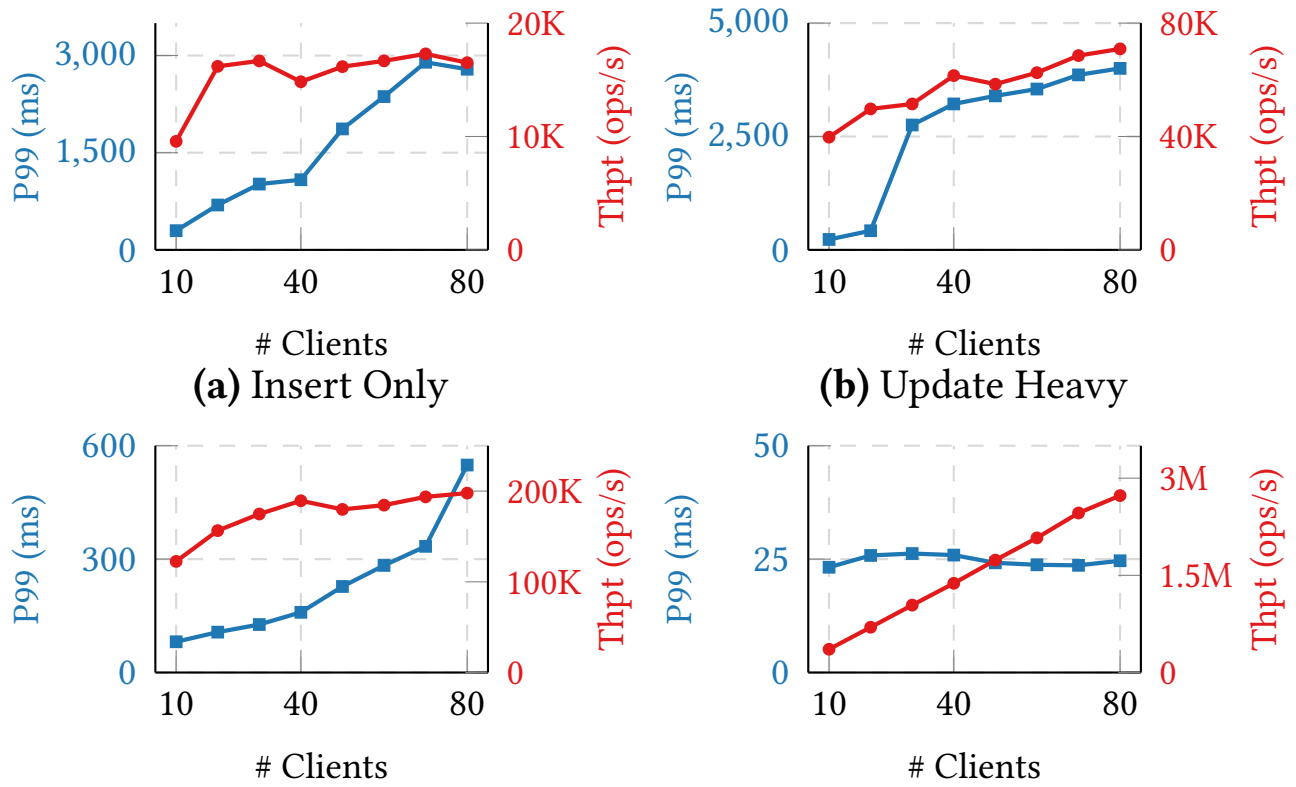
\begin{figure}[t]
    \centering
    \captionsetup{skip=4pt}
    \captionsetup[sub]{font=footnotesize,skip=1pt,belowskip=0pt,justification=centering,singlelinecheck=false}

    \subcaptionbox{Insert Only\label{fig:scalability_load}}[0.49\linewidth]{%
\begin{tikzpicture}[trim axis left,trim axis right]
            \begin{axis}[at={(0,0)},anchor=south west,
                width=\plotwidth mm,
                height=\plotheight mm,
                scale only axis,
                grid=major,
                grid style={dashed, gray!30},
                xmin=5,
                xmax=85,
                ymin=0,
                axis y line*=left, %
                ylabel={P99 (ms)},
                ylabel style={color=BlueColor, font=\scriptsize, at={(axis description cs:-0.38,0.5)}, anchor=center},
                yticklabel style={color=BlueColor, font=\scriptsize, text width=16pt, align=right},
                ymax=3500,
                ytick={0, 1500, 3000},
                xlabel={\# Clients},
                xlabel style={font=\scriptsize},
                xticklabel style={font=\scriptsize},
                xtick={10, 40, 80},
                axis x line*=bottom,
            ]
                \addplot[color=BlueColor, thick, mark=square*, mark size=0.8pt] table [
                    x=client_count, y=p99_latency_ms, col sep=comma
                ] {sections/figures/tikz_export/ssmr_scalability/scalability_load.csv};
            \end{axis}

            \begin{axis}[at={(0,0)},anchor=south west,
                width=\plotwidth mm,
                height=\plotheight mm,
                scale only axis,
                scaled ticks = false,
                scaled x ticks = false,
                xmin=5,
                xmax=85,
                ymin=0,
                axis y line*=right, %
                axis x line=none,   %
                ylabel={Thpt (ops/s)},
                ylabel style={color=RedColor, font=\scriptsize, at={(axis description cs:1.38,0.5)}, anchor=center},
                yticklabel style={color=RedColor, font=\scriptsize, text width=16pt, align=left},
                ymax=20000,
                ytick={0, 10000, 20000},
                yticklabels={0, 10K, 20K},
                xtick={10, 40, 80},
            ]
                \addplot[color=RedColor, thick, mark=*, mark size=0.8pt] table [
                    x=client_count, y=throughput_ops_s, col sep=comma
                ] {sections/figures/tikz_export/ssmr_scalability/scalability_load.csv};
                \label{plot:scalability_load_throughput}
            \end{axis}
        \pgfresetboundingbox
            \path[use as bounding box] (0,-7mm) rectangle (22mm,16mm);
        \end{tikzpicture}%
    }
    \hfill
    \subcaptionbox{Update Heavy\label{fig:scalability_workloada}}[0.49\linewidth]{%
\begin{tikzpicture}[trim axis left,trim axis right]
            \begin{axis}[at={(0,0)},anchor=south west,
                width=\plotwidth mm,
                height=\plotheight mm,
                scale only axis,
                grid=major,
                grid style={dashed, gray!30},
                xmin=5,
                xmax=85,
                ymin=0,
                axis y line*=left,
                ylabel={P99 (ms)},
                ylabel style={color=BlueColor, font=\scriptsize, at={(axis description cs:-0.38,0.5)}, anchor=center},
                yticklabel style={color=BlueColor, font=\scriptsize, text width=16pt, align=right},
                ymax=5000,
                ytick={0, 2500, 5000},
                xlabel={\# Clients},
                xlabel style={font=\scriptsize},
                xticklabel style={font=\scriptsize},
                xtick={10, 40, 80},
                axis x line*=bottom,
            ]
                \addplot[color=BlueColor, thick, mark=square*, mark size=0.8pt] table [
                    x=client_count, y=p99_latency_ms, col sep=comma
                ] {sections/figures/tikz_export/ssmr_scalability/scalability_workloada.csv};
            \end{axis}

            \begin{axis}[at={(0,0)},anchor=south west,
                width=\plotwidth mm,
                height=\plotheight mm,
                scale only axis,
                scaled ticks = false,
                scaled x ticks = false,
                xmin=5,
                xmax=85,
                ymin=0,
                axis y line*=right,
                axis x line=none,
                ylabel={Thpt (ops/s)},
                ylabel style={color=RedColor, font=\scriptsize, at={(axis description cs:1.38,0.5)}, anchor=center},
                yticklabel style={color=RedColor, font=\scriptsize, text width=16pt, align=left},
                ymax=80000,
                ytick={0, 40000, 80000},
                yticklabels={0, 40K, 80K},
                xtick={10, 40, 80},
            ]
                \addplot[color=RedColor, thick, mark=*, mark size=0.8pt] table [
                    x=client_count, y=throughput_ops_s, col sep=comma
                ] {sections/figures/tikz_export/ssmr_scalability/scalability_workloada.csv};
            \end{axis}
        \pgfresetboundingbox
            \path[use as bounding box] (0,-7mm) rectangle (22mm,16mm);
        \end{tikzpicture}%
    }\par\smallskip

    \subcaptionbox{Read Heavy\label{fig:scalability_workloadb}}[0.49\linewidth]{%
\begin{tikzpicture}[trim axis left,trim axis right]
            \begin{axis}[at={(0,0)},anchor=south west,
                width=\plotwidth mm,
                height=\plotheight mm,
                scale only axis,
                grid=major,
                grid style={dashed, gray!30},
                xmin=5,
                xmax=85,
                ymin=0,
                axis y line*=left,
                ylabel={P99 (ms)},
                ylabel style={color=BlueColor, font=\scriptsize, at={(axis description cs:-0.38,0.5)}, anchor=center},
                yticklabel style={color=BlueColor, font=\scriptsize, text width=16pt, align=right},
                ymax=600,
                ytick={0, 300, 600},
                xlabel={\# Clients},
                xlabel style={font=\scriptsize},
                xticklabel style={font=\scriptsize},
                xtick={10, 40, 80},
                axis x line*=bottom,
            ]
                \addplot[color=BlueColor, thick, mark=square*, mark size=0.8pt] table [
                    x=client_count, y=p99_latency_ms, col sep=comma
                ] {sections/figures/tikz_export/ssmr_scalability/scalability_workloadb.csv};
            \end{axis}

            \begin{axis}[at={(0,0)},anchor=south west,
                width=\plotwidth mm,
                height=\plotheight mm,
                scale only axis,
                scaled ticks = false,
                scaled x ticks = false,
                xmin=5,
                xmax=85,
                ymin=0,
                axis y line*=right,
                axis x line=none,
                ylabel={Thpt (ops/s)},
                ylabel style={color=RedColor, font=\scriptsize, at={(axis description cs:1.38,0.5)}, anchor=center},
                yticklabel style={color=RedColor, font=\scriptsize, text width=16pt, align=left},
                ymax=250000,
                ytick={0, 100000, 200000},
                yticklabels={0, 100K, 200K},
                xtick={10, 40, 80},
            ]
                \addplot[color=RedColor, thick, mark=*, mark size=0.8pt] table [
                    x=client_count, y=throughput_ops_s, col sep=comma
                ] {sections/figures/tikz_export/ssmr_scalability/scalability_workloadb.csv};
            \end{axis}
        \pgfresetboundingbox
            \path[use as bounding box] (0,-7mm) rectangle (22mm,16mm);
        \end{tikzpicture}%
    }
    \hfill
    \subcaptionbox{Read Only\label{fig:scalability_workloadc}}[0.49\linewidth]{%
\begin{tikzpicture}[trim axis left,trim axis right]
            \begin{axis}[at={(0,0)},anchor=south west,
                width=\plotwidth mm,
                height=\plotheight mm,
                scale only axis,
                grid=major,
                grid style={dashed, gray!30},
                xmin=5,
                xmax=85,
                ymin=0,
                axis y line*=left,
                ylabel={P99 (ms)},
                ylabel style={color=BlueColor, font=\scriptsize, at={(axis description cs:-0.38,0.5)}, anchor=center},
                yticklabel style={color=BlueColor, font=\scriptsize, text width=16pt, align=right},
                ymax=50,
                ytick={0, 25, 50},
                xlabel={\# Clients},
                xlabel style={font=\scriptsize},
                xticklabel style={font=\scriptsize},
                xtick={10, 40, 80},
                axis x line*=bottom,
            ]
                \addplot[color=BlueColor, thick, mark=square*, mark size=0.8pt] table [
                    x=client_count, y=p99_latency_ms, col sep=comma
                ] {sections/figures/tikz_export/ssmr_scalability/scalability_workloadc.csv};
            \end{axis}

            \begin{axis}[at={(0,0)},anchor=south west,
                width=\plotwidth mm,
                height=\plotheight mm,
                scale only axis,
                scaled ticks = false,
                scaled x ticks = false,
                xmin=5,
                xmax=85,
                ymin=0,
                axis y line*=right,
                axis x line=none,
                ylabel={Thpt (ops/s)},
                ylabel style={color=RedColor, font=\scriptsize, at={(axis description cs:1.38,0.5)}, anchor=center},
                yticklabel style={color=RedColor, font=\scriptsize, text width=16pt, align=left},
                ymax=3500000,
                ytick={0, 1500000, 3000000},
                yticklabels={0, 1.5M, 3M},
                xtick={10, 40, 80},
            ]
                \addplot[color=RedColor, thick, mark=*, mark size=0.8pt] table [
                    x=client_count, y=throughput_ops_s, col sep=comma
                ] {sections/figures/tikz_export/ssmr_scalability/scalability_workloadc.csv};
            \end{axis}
        \pgfresetboundingbox
            \path[use as bounding box] (0,-7mm) rectangle (22mm,16mm);
        \end{tikzpicture}%
    }

    \caption{Scalability of \system{} with increasing client VMs.}
    \label{fig:scalability}
\end{figure}

\smallpara{Results}
\cref{fig:scalability} shows \system{} scaling differently with the write fraction.
For read-only workloads (\cref{fig:scalability_workloadc}), throughput grows near-linearly from roughly 300K ops/s at 10 clients to about 2.7M ops/s at 80, with P99 latency stable around 25 ms.
Workloads with writes (\cref{fig:scalability_load,fig:scalability_workloada,fig:scalability_workloadb}) eventually saturate from serialization on the shared log: throughput plateaus (near 70K ops/s for Update Heavy and 17K ops/s for Insert Only) and P99 latency rises to several seconds under high contention.
By eliminating servers, \ssmr{} shifts the performance bottleneck from a central server to shared-log serialization bandwidth. Write throughput is bounded by the underlying storage's tail-segment append rate. This bottleneck differs from conventional SMR: higher-bandwidth shared logs would raise \ssmr{}'s write ceiling.

\section{Related Work}
\label{sec:related}

\textbf{Classical SMR} \ssmr{} builds on classical State Machine Replication, leveraging Lamport's formulation of distinct roles~\cite{lamport_paxos_2001}.
In \ssmr{}, both \textit{acceptor} and \textit{learner} state reside on disaggregated storage, while the \textit{proposer} runs in each client.
Together, the clients implement a single ``logical'' learner whose materialized state is durable and directly accessible through the shared storage service.

\looseness=-1 \textbf{Disaggregated SMR} relies on materializers that play an external shared log and maintain individual copies or snapshots of state, stored in memory~\cite{balakrishnan2013tango} or on local disk~\cite{balakrishnan2020virtual}.
Periodically, to speed log playback on recovery, each materializer writes its local snapshot---a deterministic function over some prefix of the shared log---to remote storage.
A fundamental property of disaggregated SMR is that a server must download a full snapshot before serving requests.
In contrast, \ssmr{} maintains a single shared copy of materialized state, so clients serve requests immediately from disaggregated storage without downloading full state.

\textbf{Serverless+Shared Logs} A sub-class of disaggregated SMR systems deals with \textit{serverless functions}, starting with Boki~\cite{jia2021boki} and successors~\cite{halfmoon, impeller, indilog, flexlog}.
Boki lets ephemeral functions materialize state by playing a fine-grained shared log (a LogBook), keeping state in local memory and periodically storing snapshots (``auxiliary data'') via the LogBook API.
In this distinction between local and remote snapshots, Boki is closer to conventional disaggregated SMR than to \ssmr{}, which keeps one materialized copy in external storage for direct use by every client.

\textbf{Databases over Shared Logs} Hyder~\cite{bernstein2011hyder}, Tango~\cite{balakrishnan2013tango}, and Delos~\cite{balakrishnan2020virtual} use a shared log to order transactions and reconstruct state.
Each server or learner, however, materializes a private copy from the remote log, so recovery entails restoring a snapshot and replaying its suffix before serving requests.
\ssmr{} instead makes reusable state shared: clients jointly maintain one materialized copy in external storage, and a new client reads that copy without first rebuilding the complete database locally.

\textbf{Log-structured Merge (LSM)} trees~\cite{o1996log} optimize write-intensive workloads by buffering writes in memory and periodically flushing to disk.
Single-node LSMs~\cite{rocksdb} are common building blocks for distributed databases~\cite{scylladb, yugabytedb, taft2020cockroachdb}.
\ssmr{} can be viewed as a \textit{distributed LSM}: instead of each node writing its own log and compacting to its own snapshot, many clients collaboratively extend a single shared log and compact it to a shared snapshot.
Viewed this way, \ssmr{} resembles LSM-style designs over disaggregated storage~\cite{nova-lsm}, but uses the structure to improve SMR availability.

\section{Conclusion}
\label{sec:con}

SMR is simplified by shared log designs on external storage, but hobbled by their reliance on failure-prone servers to materialize the log.
\ssmr{} eliminates these servers by having clients maintain a single materialized copy of state on external storage and use the shared log for both command ordering and distributed materialization.
This removes several sources of unavailability: server failures, misbehaving clients, and cold-start overheads.
Our \ssmr{} key-value store \system{} delivers strong ACID semantics with far higher availability than conventional designs, lower cost, and simpler deployment. It runs atop commodity storage such as Azure Blob, and our Redis implementation shows how a low-latency shared log improves performance.

The mechanism also clarifies which storage primitives matter.
\ssmr{} needs a durable ordered append path and indexed access to compacted state; it does not require a database service or a private snapshot for every client.
This separation lets a deployment choose storage for its latency, cost, and durability targets while preserving the client-side state-machine interface.
It also localizes recovery because a new client reads shared indexes and the short log tail rather than reconstructing the entire database before serving requests.

Evaluation marks the design's boundary: cached reads scale with clients, while writes are capped by the shared log's append rate.
\ssmr{} best fits clients near durable shared storage that value availability, rapid startup, and elastic reads over single-node latency.
More broadly, atomic shared-log operations let clients maintain reusable state without assigning materialization to a server.

\clearpage
\bibliographystyle{plain}
\bibliography{sample}

\begin{thebibliography}{10}

\bibitem{etcd}
etcd.
\newblock \url{https://etcd.io/}.

\bibitem{kraft}
{KRaft} - {Apache} {Kafka} {Without} {ZooKeeper}.
\newblock \url{https://developer.confluent.io/learn/kraft/}.

\bibitem{schedulerbadness1}
Kubernetes failure stories.
\newblock \url{https://k8s.af/}.

\bibitem{logdevice}
Logdevice.
\newblock
  \url{https://engineering.fb.com/2017/08/31/core-infra/logdevice-a-distributed-data-store-for-logs/},
  2017.

\bibitem{schedulerbadness3}
Grafana incident report \#2.
\newblock
  \url{https://grafana.com/blog/2019/07/24/how-a-production-outage-was-caused-using-kubernetes-pod-priorities/},
  2019.

\bibitem{schedulerbadness2}
Grafana incident report \#1.
\newblock
  \url{https://grafana.com/blog/2020/01/23/how-a-gcp-persistent-disk-incident-snowballed-into-a-23-hour-outage--and-taught-us-some-important-lessons/},
  2020.
\newblock [Online; accessed 11-December-2025].

\bibitem{azure-blob}
Azure blob storage.
\newblock \url{https://azure.microsoft.com/en-us/services/storage/blobs/},
  2023.
\newblock [Online; accessed 17-October-2024].

\bibitem{azsql}
Azure sql databases.
\newblock \url{https://azure.microsoft.com/en-us/products/azure-sql/database},
  2023.
\newblock [Online; accessed 17-October-2024].

\bibitem{azcloud}
Microsoft azure: Cloud computing services.
\newblock \url{https://azure.microsoft.com/en-us}, 2023.
\newblock [Online; accessed 17-October-2024].

\bibitem{rocksdb}
Rocksdb: A persistent key-value store.
\newblock \url{https://rocksdb.org/}, 2023.
\newblock [Online; accessed 17-October-2024].

\bibitem{scylladb}
Scylladb: A fast and scalable nosql database.
\newblock \url{https://www.scylladb.com/}, 2023.
\newblock [Online; accessed 17-October-2024].

\bibitem{sqlite}
Sqlite.
\newblock \url{https://sqlite.org/index.html}, 2023.
\newblock [Online; accessed 17-October-2024].

\bibitem{ycsb}
Ycsb.
\newblock \url{https://github.com/brianfrankcooper/YCSB}, 2023.
\newblock [Online; accessed 17-October-2024].

\bibitem{yugabytedb}
Yugabytedb: A cloud native distributed sql database.
\newblock \url{https://www.yugabyte.com/}, 2023.
\newblock [Online; accessed 17-October-2024].

\bibitem{corfudb}
{CorfuDB}/{CorfuDB}, August 2025.
\newblock original-date: 2013-09-19T22:02:36Z.

\bibitem{balakrishnan2024taming}
Mahesh Balakrishnan.
\newblock Taming consensus in the wild (with the shared log abstraction).
\newblock {\em ACM SIGOPS Operating Systems Review}, 58(1):1--6, 2024.

\bibitem{balakrishnan2020virtual}
Mahesh Balakrishnan, Jason Flinn, Chen Shen, Mihir Dharamshi, Ahmed Jafri, Xiao
  Shi, Santosh Ghosh, Hazem Hassan, Aaryaman Sagar, Rhed Shi, et~al.
\newblock Virtual consensus in delos.
\newblock In {\em 14th USENIX Symposium on Operating Systems Design and
  Implementation (OSDI 20)}, pages 617--632, 2020.

\bibitem{balakrishnan2012corfu}
Mahesh Balakrishnan, Dahlia Malkhi, Vijayan Prabhakaran, Ted Wobbler, Michael
  Wei, and John~D Davis.
\newblock $\{$CORFU$\}$: A shared log design for flash clusters.
\newblock In {\em 9th USENIX Symposium on Networked Systems Design and
  Implementation (NSDI 12)}, pages 1--14, 2012.

\bibitem{balakrishnan2013tango}
Mahesh Balakrishnan, Dahlia Malkhi, Ted Wobber, Ming Wu, Vijayan Prabhakaran,
  Michael Wei, John~D Davis, Sriram Rao, Tao Zou, and Aviad Zuck.
\newblock Tango: Distributed data structures over a shared log.
\newblock In {\em Proceedings of the twenty-fourth ACM symposium on operating
  systems principles}, pages 325--340, 2013.

\bibitem{balakrishnan2021log}
Mahesh Balakrishnan, Chen Shen, Ahmed Jafri, Suyog Mapara, David Geraghty,
  Jason Flinn, Vidhya Venkat, Ivailo Nedelchev, Santosh Ghosh, Mihir Dharamshi,
  et~al.
\newblock Log-structured protocols in delos.
\newblock In {\em Proceedings of the ACM SIGOPS 28th Symposium on Operating
  Systems Principles}, pages 538--552, 2021.

\bibitem{barletta2024mutiny}
Marco Barletta, Marcello Cinque, Catello Di~Martino, Zbigniew~T Kalbarczyk, and
  Ravishankar~K Iyer.
\newblock Mutiny! how does kubernetes fail, and what can we do about it?
\newblock In {\em 2024 54th Annual IEEE/IFIP International Conference on
  Dependable Systems and Networks (DSN)}, pages 1--14. IEEE, 2024.

\bibitem{berenson1995critique}
Hal Berenson, Phil Bernstein, Jim Gray, Jim Melton, Elizabeth O'Neil, and
  Patrick O'Neil.
\newblock A critique of ansi sql isolation levels.
\newblock {\em ACM SIGMOD Record}, 24(2):1--10, 1995.

\bibitem{bernstein2011hyder}
Philip~A Bernstein, Colin~W Reid, and Sudipto Das.
\newblock Hyder-a transactional record manager for shared flash.
\newblock In {\em CIDR}, volume~11, pages 9--20, 2011.

\bibitem{speclog}
Shreesha~G Bhat, Tony Hong, Xuhao Luo, Jiyu Hu, Aishwarya Ganesan, and
  Ramnatthan Alagappan.
\newblock Low $\{$End-to-End$\}$ latency atop a speculative shared log with
  $\{$Fix-Ante$\}$ ordering.
\newblock In {\em 19th USENIX Symposium on Operating Systems Design and
  Implementation (OSDI 25)}, pages 465--481, 2025.

\bibitem{burrowschubby}
Mike Burrows.
\newblock The {Chubby} lock service for loosely-coupled distributed systems.
\newblock 2006.

\bibitem{chandra2007paxos}
Tushar~D Chandra, Robert Griesemer, and Joshua Redstone.
\newblock Paxos made live: an engineering perspective.
\newblock In {\em Proceedings of the twenty-sixth annual ACM symposium on
  Principles of distributed computing}, pages 398--407, 2007.

\bibitem{corbett2013spanner}
James~C Corbett, Jeffrey Dean, Michael Epstein, Andrew Fikes, Christopher
  Frost, Jeffrey~John Furman, Sanjay Ghemawat, Andrey Gubarev, Christopher
  Heiser, Peter Hochschild, et~al.
\newblock Spanner: Google’s globally distributed database.
\newblock {\em ACM Transactions on Computer Systems (TOCS)}, 31(3):1--22, 2013.

\bibitem{ding_scalog_2020}
Cong Ding, David Chu, Evan Zhao, Xiang Li, Lorenzo Alvisi, and Robbert
  Van~Renesse.
\newblock Scalog: Seamless reconfiguration and total order in a scalable shared
  log.
\newblock In {\em 17th USENIX Symposium on Networked Systems Design and
  Implementation (NSDI 20)}, pages 325--338, 2020.

\bibitem{fischer_flp_1985}
Michael~J Fischer, Nancy~A Lynch, and Michael~S Paterson.
\newblock Impossibility of distributed consensus with one faulty process.
\newblock {\em Journal of the ACM (JACM)}, 32(2):374--382, 1985.

\bibitem{flexlog}
Dimitra Giantsidi, Emmanouil Giortamis, Nathaniel Tornow, Florin Dinu, and
  Pramod Bhatotia.
\newblock Flexlog: A shared log for stateful serverless computing.
\newblock In {\em Proceedings of the 32nd International Symposium on
  High-Performance Parallel and Distributed Computing}, pages 195--209, 2023.

\bibitem{oceanbase}
Fusheng Han, Hao Liu, Bin Chen, Debin Jia, Jianfeng Zhou, Xuwang Teng, Chuanhui
  Yang, Huafeng Xi, Wei Tian, Shuning Tao, et~al.
\newblock Palf: Replicated write-ahead logging for distributed databases.
\newblock {\em Proceedings of the VLDB Endowment}, 17(12):3745--3758, 2024.

\bibitem{tidb}
Dongxu Huang, Qi~Liu, Qiu Cui, Zhuhe Fang, Xiaoyu Ma, Fei Xu, Li~Shen, Liu
  Tang, Yuxing Zhou, Menglong Huang, et~al.
\newblock Tidb: a raft-based htap database.
\newblock {\em Proceedings of the VLDB Endowment}, 13(12):3072--3084, 2020.

\bibitem{nova-lsm}
Haoyu Huang and Shahram Ghandeharizadeh.
\newblock Nova-lsm: a distributed, component-based lsm-tree key-value store.
\newblock In {\em Proceedings of the 2021 International Conference on
  Management of Data}, pages 749--763, 2021.

\bibitem{hunt_zookeeper_2010}
Patrick Hunt, Mahadev Konar, Flavio~P. Junqueira, and Benjamin Reed.
\newblock {ZooKeeper}: wait-free coordination for internet-scale systems.
\newblock In {\em Proceedings of the 2010 {USENIX} conference on {USENIX}
  annual technical conference}, {USENIXATC}'10, page~11, USA, 2010. USENIX
  Association.

\bibitem{jia2021boki}
Zhipeng Jia and Emmett Witchel.
\newblock Boki: Stateful serverless computing with shared logs.
\newblock In {\em Proceedings of the ACM SIGOPS 28th Symposium on Operating
  Systems Principles}, pages 691--707, 2021.

\bibitem{kreps_kafka_nodate}
Jay Kreps, Neha Narkhede, and Jun Rao.
\newblock Kafka: a {Distributed} {Messaging} {System} for {Log} {Processing}.

\bibitem{lamport_time_1978}
Leslie Lamport.
\newblock Time, {Clocks}, and the {Ordering} of {Events} in a {Distributed}
  {System}.
\newblock 21(7), 1978.

\bibitem{lamport_paxos_2001}
Leslie Lamport.
\newblock Paxos {Made} {Simple}.
\newblock {\em ACM SIGACT News (Distributed Computing Column) 32, 4 (Whole
  Number 121, December 2001)}, pages 51--58, December 2001.

\bibitem{lockerman2018fuzzylog}
Joshua Lockerman, Jose~M Faleiro, Juno Kim, Soham Sankaran, Daniel~J Abadi,
  James Aspnes, Siddhartha Sen, and Mahesh Balakrishnan.
\newblock The fuzzylog: A partially ordered shared log.
\newblock In {\em OSDI}, pages 357--372, 2018.

\bibitem{lazylog}
Xuhao Luo, Shreesha~G Bhat, Jiyu Hu, Ramnatthan Alagappan, and Aishwarya
  Ganesan.
\newblock Lazylog: A new shared log abstraction for low-latency applications.
\newblock In {\em Proceedings of the ACM SIGOPS 30th Symposium on Operating
  Systems Principles}, pages 296--312, 2024.

\bibitem{zippydb}
Sarang Masti.
\newblock How we built a general purpose key value store for {Facebook} with
  {ZippyDB}, August 2021.

\bibitem{nawab2015chariots}
Faisal Nawab, Vaibhav Arora, Divyakant Agrawal, and Amr El~Abbadi.
\newblock Chariots: A scalable shared log for data management in
  multi-datacenter cloud environments.
\newblock In {\em EDBT}, pages 13--24, 2015.

\bibitem{raft}
Diego Ongaro and John Ousterhout.
\newblock In search of an understandable consensus algorithm.
\newblock In {\em 2014 USENIX annual technical conference (USENIX ATC 14)},
  pages 305--319, 2014.

\bibitem{o1996log}
Patrick O’Neil, Edward Cheng, Dieter Gawlick, and Elizabeth O’Neil.
\newblock The log-structured merge-tree (lsm-tree).
\newblock {\em Acta informatica}, 33(4):351--385, 1996.

\bibitem{postgresql}
{PostgreSQL Global Development Group}.
\newblock {PostgreSQL}.
\newblock \url{https://www.postgresql.org/}, April 2025.

\bibitem{halfmoon}
Sheng Qi, Xuanzhe Liu, and Xin Jin.
\newblock Halfmoon: Log-optimal fault-tolerant stateful serverless computing.
\newblock In {\em Proceedings of the 29th Symposium on Operating Systems
  Principles}, pages 314--330, 2023.

\bibitem{myraft}
Anirban Rahut, Vinaykumar Bhat, Abhinav Sharma, Yichen Shen, Bartlomiej Pelc,
  Chi Li, Ahsanul Haque, Yash Botadra, Xi~Wang, Michael Percy, et~al.
\newblock Myraft: High availability in mysql using raft.
\newblock In {\em EDBT}, pages 743--752, 2024.

\bibitem{schneider_implementing_1990}
Fred~B. Schneider.
\newblock Implementing fault-tolerant services using the state machine
  approach: a tutorial.
\newblock {\em ACM Computing Surveys}, 22(4):299--319, December 1990.

\bibitem{taft2020cockroachdb}
Rebecca Taft, Irfan Sharif, Andrei Matei, Nathan VanBenschoten, Jordan Lewis,
  Tobias Grieger, Kai Niemi, Andy Woods, Anne Birzin, Raphael Poss, et~al.
\newblock Cockroachdb: The resilient geo-distributed sql database.
\newblock In {\em Proceedings of the 2020 ACM SIGMOD international conference
  on management of data}, pages 1493--1509, 2020.

\bibitem{memorydb}
Yacine Taleb, Kevin McGehee, Nan Yan, Shawn Wang, Stefan~C M{\"u}ller, and
  Allen Samuels.
\newblock Amazon memorydb: A fast and durable memory-first cloud database.
\newblock In {\em Companion of the 2024 International Conference on Management
  of Data}, pages 309--320, 2024.

\bibitem{verbitski2017amazon}
Alexandre Verbitski, Anurag Gupta, Debanjan Saha, Murali Brahmadesam, Kamal
  Gupta, Raman Mittal, Sailesh Krishnamurthy, Sandor Maurice, Tengiz
  Kharatishvili, and Xiaofeng Bao.
\newblock Amazon aurora: Design considerations for high throughput cloud-native
  relational databases.
\newblock In {\em Proceedings of the 2017 ACM International Conference on
  Management of Data}, pages 1041--1052, 2017.

\bibitem{wei2017vcorfu}
Michael Wei, Amy Tai, Christopher~J Rossbach, Ittai Abraham, Maithem Munshed,
  Medhavi Dhawan, Jim Stabile, Udi Wieder, Scott Fritchie, Steven Swanson,
  et~al.
\newblock $\{$vCorfu$\}$: A $\{$Cloud-Scale$\}$ object store on a shared log.
\newblock In {\em 14th USENIX Symposium on Networked Systems Design and
  Implementation (NSDI 17)}, pages 35--49, 2017.

\bibitem{indilog}
Maximilian Wiesholler, Florin Dinu, Javier Picorel, and Pramod Bhatotia.
\newblock Indilog: Bridging scalability and performance in stateful serverless
  computing with shared logs.
\newblock In {\em Proceedings of the 17th ACM International Systems and Storage
  Conference}, pages 1--13, 2024.

\bibitem{upgradefailures}
Yongle Zhang, Junwen Yang, Zhuqi Jin, Utsav Sethi, Kirk Rodrigues, Shan Lu, and
  Ding Yuan.
\newblock Understanding and detecting software upgrade failures in distributed
  systems.
\newblock In {\em Proceedings of the ACM SIGOPS 28th Symposium on Operating
  Systems Principles}, pages 116--131, 2021.

\bibitem{impeller}
Zhiting Zhu, Zhipeng Jia, Newton Ni, Dixin Tang, and Emmett Witchel.
\newblock Impeller: Stream processing on shared logs.
\newblock In {\em Proceedings of the Twentieth European Conference on Computer
  Systems}, pages 637--653, 2025.

\end{thebibliography}

\end{document}